\documentclass[aps,preprint,superscriptaddress]{revtex4}
\pdfoutput=1
\usepackage{amsmath,amssymb,amsfonts}
\usepackage{bm}

\usepackage{braket,bm,bbm,setspace}
\usepackage[normalem]{ulem} 
\usepackage{physics}
\usepackage[makeroom]{cancel}
\usepackage{graphicx,overpic,mathtools}
\usepackage{amsthm,amsmath,amssymb,hyperref,mathrsfs}
\usepackage{braket,bm,bbm,setspace}
\usepackage[normalem]{ulem} 
\usepackage{physics}
\usepackage[makeroom]{cancel}
\hypersetup{
    colorlinks=false,
    pdfborder={0 0 0},
}
\usepackage[usenames,dvipsnames]{xcolor}
\usepackage{graphicx}
\usepackage{bbold}
\usepackage{slashed}
\usepackage{color}
\usepackage{hyperref}  
\hypersetup{colorlinks=true, linkcolor=blue, citecolor=red}  

\newcommand{\ii}{\textrm {i}}
\newcommand{\de}[1]{\partial_{#1}}

\newcommand{\br}[1]{\left( #1 \right)}

\newcommand{\Hyper}[1]{ {}_2 \mathcal F_1 \br{#1}}

\usepackage[T1]{fontenc} 

\begin{document}
\title{Vacuum polarization current in presence of intense Sauter field}

\author{Deepak Sah\footnote{Corresponding author.\\E-mail address: deepakk@rrcat.gov.in \& dsah129@gmail.com (Deepak Sah).}}
\author{Manoranjan P. Singh}
\affiliation{Homi Bhabha National Institute, Training School Complex, Anushakti Nagar, Mumbai 400094, India}
\affiliation{Theory and Simulations Lab, Raja Ramanna Centre for Advanced
Technology,  Indore-452013, India}


%
%
%
%
\begin{abstract}
The quantum vacuum becomes unstable under the action of an external field, leading to the spontaneous creation of charged particle-antiparticle pairs—a famous phenomenon known as Sauter-Schwinger pair production. It is often noted and emphasized that the time-dependent number of created particles, as expressed through Bogoliubov transformations within the canonical quantization approach, lacks physical meaning until the external field is switched off.  In this context, it is useful to explore dynamical physical quantities that remain well-defined not only at asymptotic times but also at intermediate times, such as the vacuum polarization current. Investigating the generic features of this observable may provide insights into the intermediate-time behavior of dynamical variables in the system’s evolution equations. For that, we consider the creation of particle-antiparticle pairs in a spatially homogeneous, time-dependent, intense Sauter field. Focusing on scalar QED in (1+1) dimensions, we examine the properties of the particle number and the correlation function. Specifically, we analyze the real and imaginary parts of the correlation function, which can be linked to vacuum polarization effects.The vacuum polarization current in the presence of an intense laser pulse is calculated numerically. Finally, we explore its dynamical behavior and find that it correlates with the real part of the correlation function. Initially, the current exhibits a sign change and gradually decreases. However, unlike the time evolution of the particle number, the current does not reach a constant asymptotic value. Instead, for $t \gg \tau$, it exhibits nearly undamped oscillations, a distinctive feature of the scalar particle system, where the current oscillates strongly around zero. Furthermore, we discuss the uniqueness of the vacuum polarization current in the adiabatic basis, comparing two different choices related to the reference  mode functions. Notably, we find that the current remains independent of the choice of basis.
\end{abstract}
%
%

\maketitle
\tableofcontents
\section{Introduction}
\label{intro}
The production of particle-antiparticle pairs from the vacuum under the influence of strong electromagnetic or gravitational fields is a fundamental quantum phenomenon. Initially explored in the late 1920s and early 1930s within relativistic quantum mechanics by Klein \cite{Klein1927,Klein:1929zz}and Sauter\cite{Sauter:1931zz,Sauter:1932gsa}, a complete theoretical framework emerged only with the advent of quantum field theory (QFT). In this framework, pair production arises due to the interaction of a quantum matter field with a time-dependent external field. This phenomenon manifests in various contexts, such as cosmological particle production, which is driven by the evolution of spacetime\cite{Parker:1969au,Gibbons:1977mu}, and the Schwinger effect, where a strong electric field induces vacuum decay \cite{Heisenberg:1936nmg,Schwinger:1951nm,Schwinger:1958wrk}.
The Schwinger effect is a central prediction of strong-field quantum electrodynamics (QED), yet its direct experimental verification remains challenging due to the exponential suppression of pair production at field strengths below the critical value $E_c  = m^2/e\simeq 1.32 \times  10^{18} V/m$ \cite{Sauter:1931zz}  (see \cite{Ringwald:2001ib} for details on the laser field parameters). However, recent advances in high-power laser technology, including facilities such as European Extreme Light Infrastructure for Nuclear Physics (ELI-NP)\cite{ELI, Heinzl:2009bmy, Dunne:2008kc}, X-ray free electron laser (XFEL) at DESY \cite{DESY}, Hamburg using self amplified spontaneous emission (SASE) principle\cite{TESLA:2000tvm}, Center for Relativistic Laser Science (CoReLS)\cite{CoReLS}, are bringing strong-field QED effects, including the Schwinger effect, within experimental reach\cite{Burke:1997ew}. These ultra-intense laser pulses have significantly impacted various fields, including relativistic nonlinear optics, high-field QED, high-energy particle physics, and laboratory astrophysics see ref.\cite{Fedotov:2022ely,Yu2023-ae}.
\par
Beyond high-energy physics, vacuum pair production has become an experimentally accessible phenomenon in condensed matter systems, particularly in graphene research \cite{CastroNeto:2007fxn, Vozmediano:2010zz,Schmitt:2022pkd,Villalba-Chavez:2022flx}. The low-energy excitations in a monolayer of graphene under an external electromagnetic field can be effectively described by the Dirac equation in (2+1) dimensions\cite{Allor:2007ei,Semenoff:1984dq,Smolyansky:2019dqd,Klimchitskaya:2013fpa,Lewkowicz2011SignatureOT}. Recent experiments have observed nonlinear current-voltage characteristics in graphene devices, interpreted in terms of particle-antiparticle pair creation\cite{PhysRevB.82.045416}. This has motivated further theoretical studies exploring the possibility of achieving supercritical field strengths and understanding the fundamental limits of pair production in condensed matter analogues \cite{Smolyansky:2020rbh, Panferov:2019stq, Smolyansky:2017mvh,Schmitt:2022pkd}.
\par
Experimental proposals involving ultra-intense lasers have positioned vacuum pair creation as one of the first non-perturbative QFT effects to be tested in a controlled laboratory setting \cite{Yakimenko:2018kih,DiPiazza:2011tq,Fedotov:2022ely}. This provides a unique opportunity to probe the non-perturbative regime of QFT beyond the standard perturbative expansion techniques . From a theoretical perspective, pair production is deeply connected to the concept of the quantum vacuum \cite{Ford:2021syk,Alvarez-Dominguez:2023zsk,Blinne:2018nbd}. In the canonical quantization of matter fields, vacuum states are usually defined by preserving the symmetries of the classical theory. In Minkowski spacetime, Poincaré symmetry uniquely determines the annihilation and creation operators, leading to a well-defined vacuum. However, in the presence of a time-dependent external field, time translation symmetry is broken, making the definition of particles ambiguous \cite{Cortez:2021njc, Cortez:2019orm}. Different choices of vacuum states lead to different quantum theories, each with distinct notions of particles. Various vacuum selection criteria have been proposed, including adiabatic vacua \cite{Lueders:1990np}, Hamiltonian diagonalization \cite{ElizagaNavascues:2019itm,Fahn:2018ahm}, and methods based on minimizing the renormalized stress-energy tensor \cite{Agullo:2014ica, Yamada:2021kqw}.
An evolving external field can dynamically modify the system’s vacuum state, leading to the creation of particle-antiparticle pairs. The time dependent number of created pairs,$\mathcal{N}(t)$ depends on the vacuum states chosen for comparison \cite{Dabrowski:2014ica}. At asymptotic times, once the external field has settled, the particle concept becomes well-defined \cite{Taya:2020dco, Dabrowski:2016tsx}. However, during the field’s evolution, particle definitions remain ambiguous, necessitating alternative approaches such as quasi-particle descriptions \cite{Blaschke:2013ip, Gregori:2010uf}. This has led to question: is there a well-defined observable that characterizes pair production dynamics across all time not only asymptotic time.
A promising candidate for such an observable is the vacuum polarization current density, which is physically meaning full over all time.
Motivated by this, we analyze the behavior of the vacuum polarization current density. In present work we are interested in generic features of this observable  to gain insights into the intermediate-time behavior of the dynamical variable in the system's time evolution equation.
\begin{figure}
    \centering
    \includegraphics[width=0.5\linewidth]{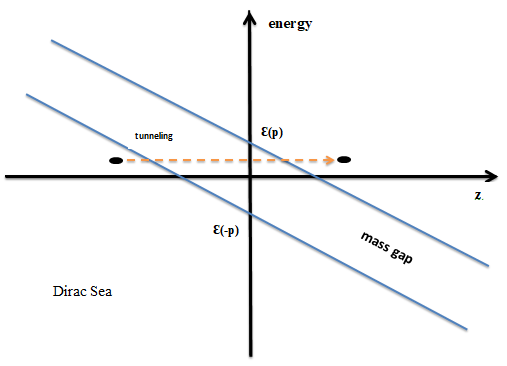}
    \caption{Schematic picture of a tunneling process}
    \label{Tunnel}
\end{figure}
The polarization current can be understood in terms of the variation of an electric dipole within the canonical quantization framework. Consider a particle-antiparticle pair created with momenta \(\bm{p}\) and \(-\bm{p}\). When these particles become on-shell, their separation is given by \( 2 \frac{\varepsilon(\bm{p})}{eE} \), which can be interpreted as forming an electric dipole of magnitude \( 2e \frac{\varepsilon(\bm{p})}{eE} \) (see Fig.\ref{Tunnel}).  
Since the polarization current arises from the variation of an electric dipole, its creation rate is given by \( \frac{d}{dt} \mathcal{N}(\bm{p},t) \). Thus, the total variation rate of the electric dipole, corresponding to the polarization current, is  $J^{z}_{\text{pol}}(t) = \int [d\bm{p}] \, 2e \frac{\varepsilon(\bm{p})}{eE} \frac{d}{dt} \mathcal{N}(\bm{p},t).$
In the quantum kinetic formalism, the rate of particle creation is linked to the correlation function of single-particle states as  
$\frac{d}{dt} \mathcal{N}(\bm{p},t) \propto \mathcal{Re}(\bra{0}  \hat{a}^\dagger (p,t) \hat{b}^\dagger (-p,t) \ket{0}),$ demonstrating its connection to the polarization current.
To analyze this effect, we consider scalar quantum electrodynamics (QED) in a spatially homogeneous, time-dependent electric field. 
Specifically, we consider a Sauter-type electric field profile, which allows for analytical solutions to the 
particle number, or a distribution function \cite{Narozhnyi:1970uv,Gavrilov:1996pz,Gelis:2015kya,Hebenstreit:2010vz} and pair production dynamics \cite{Blaschke:2008wf,Banerjee:2018azr}. First, we revisit the standard expression found in the literature for the temporal evolution of particle number, considering arbitrary mode selections for quantization. This issue is addressed using Bogoliubov transformations within the canonical quantization approach. The evolution of the number of created particles exhibits distinct dynamical stages, as shown in Ref. \cite{Smolyansky:2016gmp} for spin-1/2 fermions, with similar behavior observed for scalar particles. Natural choices of reference mode functions yield results equivalent to those obtained via the quantum kinetic formalism \cite{Schmidt:1998vi,Fedotov:2010ue}.
However, the evolution  of the correlation function, $\mathcal{C}(\bm{p},t)$ which represent the vacuum pair creation and annihilation effects, has not been studied so far, to the best of our knowledge.  The  real part of the correlation function  $u(\bm{p},t)$ gives vacuum polarization, where as the imaginary part  $v(\bm{p},t)$  represent the corresponding counter term.
These quantities are directly related to the polarization current density and provide deeper insight into the underlying physics of pair production.
We  investigate the  polarization current density and analyze its interrelation with the correlation functions. 
\newline
The paper is structured as follows. In Sec. II, we provide a detailed theoretical formulation of the problem, following the derivations in \cite{Kluger:1998bm,Grib:1972te}, but presented here for completeness. Sec. III introduces the generalized expression for the number of created particles and particle correlation functions, incorporating arbitrary mode choices for quantization via the Bogoliubov transformation. In Sec. IV, we derive the time-dependent particle correlation function in the presence of an intense Sauter field, while Sec. V discusses its dynamical features. Additionally, in Sec. VI, we examine the basis dependence of the polarization current. Finally, the paper is summarized in the concluding section.
\newline 
Throughout the paper, we use natural units and set $ \hslash = c = m =|e|= 1 $, the electric charge $e < 0$, and express all variables in terms of the electron mass unit.
\section{Theory}
\label{Canonical}
 Let us consider a charged scalar field $\Phi$ coupled to an electromagnetic field specified by the
four-vector potential $A_{\mu}$. Its Lagrangian density is given by 
\begin{align}
      \mathcal{L} &= (\partial_\mu + \ii e A_{\mu}) \Phi^* (\partial^\mu + \ii e A^\mu ) \Phi - m^2 \Phi^* \Phi - \frac{1}{4}  F_{\mu \nu } F^{\mu \nu}.
      \label{eqn01}
\end{align}
where $e$ is the electric charge of the charged field, $m$ is its mass,$F_{\mu \nu } = \partial_\mu  A_{\nu} - \partial_\nu A_{\mu}$  is the electromagnetic field -strength  tensor, and the symbol $*$ denotes complex conjugation.

By use of the Euler-Lagrange equation we are able to derive the equation of motion for
the field
\begin{align}
     [ (\partial^\mu + \ii e  A^\mu  ) ( \partial_\mu + \ii e A_{\mu}) + m^2] \Phi (t,x) &= 0.
     \label{eqn02}
\end{align}
In this work, we will assume that the electric field is spatially homogenoeus  but time dependent.  We will use the temporal gauge to express the $A_\mu  = (0,0,0,A(t))$
and so that the electric field is 
\begin{align}
    E(t) = -\dot{A}(t) \hat{z}.
    \label{eqn03}
\end{align}
As is customary when using the canonical formalism, we require that the canonical equal-time commutation relations  for the fields are satisfied \cite{Peskin}.
\begin{align}
    [\Phi(x, t), \Pi(x, t)] &= \ii \delta(x - x'). 
    \label{eqn4}
\end{align}
The charged field \(\Phi(x, t)\) and the conjugate momentum field \(\Pi(x, t)\), along with their complex conjugates, are expanded into Fourier modes.
\begin{align}
    \Phi(\bm{x}, t) &= \int \phi(p, t) e^{i \bm{p} \cdot \bm{x}} [dp],
\label{eqn5}
\end{align}
\begin{align}
    \Pi(\bm{x}, t) = \int \pi(p, t) e^{-i \bm{p} \cdot  \bm{x}} \, [dp],
\label{eqn6}
\end{align}
where we used \(\pi(\bm{p}, t) = \dot{\phi}^*(\bm{p}, t)\) and \(\pi^*(\bm{p}, t) = \dot{\phi}(\bm{p}, t)\) \cite{Peskin}. Here, 
$
[d\bm{p}] \equiv \frac{d^3\bm{p}}{(2\pi)^3}.
$
\newline
Using the equal-time commutation relations  and inserting the inverse Fourier transformations
\begin{align}
    \phi(\bm{p
    }, t) = \int \Phi(\bm{x}, t) e^{-i p \cdot \bm{x}} \, d^3x 
    \label{eqn7}
\end{align}
\begin{align}
   \pi(\bm{p}, t) = \int \Pi(\bm{x}, t) e^{i p \cdot \bm{x}} \, d^3x 
   \label{eqn8}
\end{align}
and corresponding equal-time commutation relations for the transformed fields:
\begin{align}
     [\phi(\bm{p}, t), \pi(\bm{p'}, t)] = i(2\pi)^3 \delta(\bm{p} - \bm{p'}) ,
    \label{eqn9}
\end{align}
\begin{align}
    [\phi^*(\bm{p}, t), \pi^*(\bm{p'}, t)] = i(2\pi)^3 \delta(\bm{p} - \bm{p'}) .
    \label{eqn10}
\end{align}

\subsection{ Canonical Quantization approach}

Let us consider the particular solution of the time-dependent  harmonic oscillator differential equation \eqref{eqn02} described by  
$f(\bm{p},t),$ and there exists a unique  complex coefficient $a(\bm{p})$ and $b^*(\bm{p})$ such that any other  complex mode solution $\phi(\bm{p}, t)$ and its canonically conjugate momentum  $\pi(\bm{p}, t) $ can be uniquely written as 
\begin{align}
     \begin{pmatrix}
\phi(\bm{p}, t)  \\
\pi(\bm{p}, t)
\end{pmatrix} & = \mathbf{M}_1   \begin{pmatrix}
a(\bm{p})  \\
b^*( \bm{p})
\end{pmatrix}
\label{eqn11}
\end{align}

\begin{align}
\mathbf{M}_1&= \begin{pmatrix}
f(\bm{p}, t) & f^*(\bm{p}, t) \\
\dot{f}(\bm{p}, t) & \dot{f}^*(\bm{p}, t)
\end{pmatrix}
\label{eqn12}
\end{align}
The coefficient $a(\bm{p})$ and $b^*(\bm{p})$ of this
linear combination are called annihilation and creation
variables, respectively. 
The time dependence in this basis is governed by the complex mode function \( f(\bm{p}, t) \), which satisfies the equation of motion,
\begin{eqnarray}
       \Bigg( \frac{d^2}{d t^2}  + \omega(\bm{p},t)  \Bigg) f(\bm{p},t) &=0,
       \label{eom}
\end{eqnarray}
where the time dependent frequency $\omega(\bm{p},t)$ is given by
\begin{align}
    \omega(\bm{p},t) &= \sqrt{ (\bm{p} - e \bm{A}(t))^2 + m^2}.
    \label{omega}
\end{align}
Thus, the decomposition \eqref{eqn12} depends on the particular choice of the solution $f(\bm{p}, t)$.
\par
In a more general context, we can express the solution $\phi(\bm{p}, t)$and its conjugate momentum $\pi(\bm{p}, t)$ in terms of complex functions $\chi (\bm{p},t)$  and $ \psi (\bm{p},t) $, respectively.
In canonical studies of quantum theory dynamics (or unitary dynamics), these functions are typically chosen as \cite{Garay:2019pjo,Cortez:2021njc}:

\begin{align}
    \begin{pmatrix}
    \phi(\bm{p},t) \\
    \pi(\bm{p},t)
\end{pmatrix}
= \mathbf{M}_2 
\begin{pmatrix}
    a(\bm{p},t) \\
    b^*(\bm{p},t)
\end{pmatrix},
\label{eqn13}
\end{align}

where the matrix $\mathbf{M}_2$ is given by:
\begin{align}
   \mathbf{M}_2 =
\begin{pmatrix}
    \chi(\bm{p},t) & \chi^*(\bm{p},t) \\
    \psi(\bm{p},t) & \psi^*(\bm{p},t)
    \end{pmatrix}.
\label{eqn14}
\end{align}

It is important to note that $ \chi(\bm{p},t)$ does not necessarily solve the time-dependent harmonic oscillator equation \eqref{eqn02}. If $ \chi(\bm{p},t)$ satisfies the harmonic oscillator equation, the annihilation and creation variables  $a(\bm{p},t) $ and $ b^*(\bm{p},t)$   will be time-independent. However, if this condition is not met, the variables $a(\bm{p},t) $ and  $ b^*(\bm{p},t)$ must incorporate the necessary time dependence to account for the time dependence of $\chi(\bm{p},t)$and $\psi(\bm{p},t) $, ensuring that the field solution $\phi(\bm{p},t) $ remains valid\cite{Habib:1999cs,Kluger:1998bm}.
Thus, one can obtain the normalization condition:
\begin{equation}
    \chi(\bm{p},t)\psi^*(\bm{p},t) - \chi^*(\bm{p},t)\psi(\bm{p},t) = i.
    \label{eqn16}
\end{equation}


\par
In canonical quantization, the annihilation and creation variables \( a(\bm{p},t) \) and \( b^*(\bm{p},t) \) are promoted to operators \( \hat{a}(\bm{p},t) \) and \( \hat{b}^\dagger(\bm{p},t) \), which act on the Fock space. The vacuum state \( |0\rangle_t \) is annihilated by the operator \( \hat{a}(\bm{p},t) \) for all values of \( \bm{p}\).
This leads to the understanding that a classical theory can correspond to infinitely many quantum theories, each defined by a different choice of functions \( (\chi(\bm{p},t), \psi(\bm{p},t)) \). These choices reflect the freedom in selecting the complex structure used for quantization, as encoded in these functions\cite{Garay:2019pjo}.
In the literature, the quantization procedure enforces a unitarity condition on the field equations, which is intrinsically linked to the symmetry of the classical system. In Minkowski spacetime, the Poincaré symmetry uniquely determines the functions $\chi(\bm{p},t) $  and $\psi(\bm{p},t)) $ as plane waves with frequency $\omega(\bm{p},t)$. When an external field varies slowly over time, this framework can be extended using a WKB-like solution, allowing $\chi(\bm{p},t) $  and $\psi(\bm{p},t)) $ to approximate the Minkowski case in the limit where the frequency remains nearly constant \cite{Parker:1969au,Winitzki:2005rw,Enomoto:2020xlf,Dumlu:2010ua,Yamada:2021kqw}.

\subsection{Parametrization in the Mode Function}

One can consider the $\chi(\bm{p},t)$ as WKB-like ansatz \cite{Kluger:1998bm,Taya:2020dco,Hashiba:2022bzi}
\begin{align}
     \chi (\bm{p},t) &= \frac{1}{\sqrt{2 \mathcal{W} (\bm{p},t)}} e^{-i \Theta (\bm{p},t)},
     \label{eqn17}
\end{align}
that have two arbitrary real functions, $\mathcal{W} (\bm{p},t) > 0$  and $\Theta (\bm{p},t)$, which represent its modulus and phase, respectively and need to be specified. 
Furthermore, it is straightforward to verify that the normalization condition reduces the non-uniqueness  in the choice of the complex function $\psi(\bm{p},t)$

\begin{align}
     \psi(\bm{p},t) &= - \sqrt{\frac{\mathcal{W}(\bm{p},t)}{2}} \left[ i + \mathcal{Y} (\bm{p},t)\right] e^{-i \Theta(\bm{p},t)}.
     \label{eqn18}
\end{align}
In some cases, specific families of functions
$ ( \chi (\bm{p},t),\psi(\bm{p},t))$
 are of particular interest. In this particular case, one can insist that  $\chi (\bm{p},t)$ is a solution to the time-dependent harmonic oscillator equation \eqref{eom} with time-dependent frequency $ \mathcal{W}(\bm{p},t)$, it is not only necessary for the normalization condition to be satisfied, but also for
 $\psi(\bm{p},t) = \dot{\chi}(\bm{p},t)$ \cite{Garay:2019pjo,Barry:1989zz,Yamada:2021kqw}. This condition uniquely determines \( \dot{\Theta}(\bm{p},t) \) and \( \mathcal{Y}(\bm{p},t) \) as functions of \(  \mathcal{W}(\bm{p},t) \), according to the relations:

 \begin{align}
 \mathcal{W}(\bm{p},t)^2 &= \omega( \bm{p},t)^2 - \frac{1}{2} \left[ \frac{\ddot{\mathcal{W}}(\bm{p},t)}{\mathcal{W}(\bm{p},t)} - \frac{3}{2} \frac{\dot{ \mathcal{W}}(\bm{p},t)^2}{\mathcal{W}(\bm{p},t)^2} \right],
 \label{eqn19}
 \end{align}
\begin{align}
    \dot{\Theta}(\bm{p},t) &=  \mathcal{W}(\bm{p},t), \quad \mathcal{Y} (\bm{p},t) = \frac{\dot{\mathcal{W}}(\bm{p},t)}{2 \mathcal{W}(\bm{p},t)^2}.
    \label{eqn20}
\end{align}

Thus, the freedom in choosing the pair \( (\chi( \bm{p},t),\psi( \bm{p},t)) \), when imposing that $\chi( \bm{p},t)$ is a specific normalized solution to the equation \eqref{eom}, is encoded in the initial conditions $\mathcal{W}(\bm{p},t_0) $, \( \dot{\mathcal{W}}(\bm{p},t_0) \), and \( \Theta(\bm{p},t_0)\) at some initial time \( t_0 \).

Alternatively, one may require that $\chi( \bm{p},t)$ is an approximate solution to the equation of motion. In this case, the equations for $\mathcal{W}(\bm{p},t)$and $\mathcal{Y} (\bm{p},t)$ hold approximately. For instance, when the time-dependent frequency \( \omega(\bm{p},t) \) varies slowly, the adiabatic approximation is commonly employed \cite{Kluger:1998bm}.
It can be viewed as a generalization of the WKB approximation, with the WKB approximation representing the zeroth order of the adiabatic expansion.
\begin{align}
    \mathcal{W}^{(0)}(\bm{p},t)  = \omega(\bm{p},t), \Theta^{(0)}(\bm{p},t) =\int_{t_0}^t dt' \, \omega(\bm{p},t').
\label{eqn21}
\end{align}
For \( \mathcal{Y}^{(0)}(\bm{p},t) \), there are two possible choices, corresponding to different adiabatic orders (i.e., the number of time derivatives of \( \omega(\bm{p},t) \)). The first choice, \( \mathcal{Y}^{(0)}(\bm{p},t)  = \frac{\dot{\omega}(\bm{p},t)}{2 \omega(\bm{p},t)^2} \), is frequently used in quantum field theory in curved spacetime \cite{Yamada:2021kqw,Birrell,Dabrowski:2014ica}. This corresponds to approximating exact modes (called zeroth-order adiabatic modes) and their derivatives up to second-adiabatic order. The second choice, \( 
\Tilde{\mathcal{Y}}^{(0)}(\bm{p},t)  = 0 \), is often employed in studies of  quantum kinetic equation for particle production in the Schwinger mechanism  and approximates exact adiabatic modes up to first-adiabatic order \cite{Kluger:1991ib,Smolyansky:2019dqd,Smolyansky:1997fc,Kluger:1992gb,Dumlu:2010ua}. 
The implications of these two selections will be discussed in section \ref{invariance}. 
\newline

In general, the \( n \)-th adiabatic approximation is obtained by recursively incorporating the previous order into the equation \eqref{eqn19} for \( \mathcal{W} (\bm{p},t)\). The corresponding exact mode \( f^{(n)}(\bm{p},t) \), determined by the initial conditions \( f^{(n)}(\bm{p},t_0)= \chi^{(n)}(\bm{p},t_0) \) and \( \dot{f}^{(n)}(\bm{p},t_0)= \psi^{(n)}(\bm{p},t_0) \), is referred to as the \( n \)-th-order adiabatic mode.  However, in general, \( (\chi_{\mathbf{k}}(t), \psi_{\mathbf{k}}(t)) \) are not required to be exact solutions of the equation of motion. Various alternative choices have been explored in the literature, including functions that diagonalize the Hamiltonian at large wave numbers\cite{Cortez:2021njc} or those that minimize oscillations in the number of created particles over time \cite{Dabrowski:2016tsx,Ilderton:2021zej}.




\section{Dynamical quantity}
\label{key qunatity}
\subsection{Number of Created Particles} 
\label{sec_N(t)}
In this section, we discuss the computation of the time-dependent particle number using the canonical  transformation approach, as widely explored in references~\cite{Kluger:1991ib,Smolyansky:1997fc,Kluger:1992gb,Blaschke:2017igl}. Our goal is to determine how the particle number evolves within the vacuum states \(|0\rangle_t\) and compare them to the vacuum state of another quantum theory chosen as a reference.  
To analyze this evolution, we examine the behavior of the functions \(\chi(\bm{p},t)\) and \(\psi(\bm{p},t)\) over time, which describe the quantization process where particles are created or annihilated as time progresses. To facilitate this comparison, we first select a reference vacuum and fix a complex basis $( f(\bm{p},t), f^*(\bm{p},t) )$ for the space of solutions to the harmonic oscillator equation  \eqref{eqn13}  from which we define the associated creation and annihilation operators \( a(\bm{p}) \) and \( b^\dagger (\bm{p}) \), which remain time-independent.  The reference vacuum, denoted by \(\ket{0}\), is defined as the state annihilated by  \(\hat{a}(\bm{p})\) operator. This formulation allows us to study the evolution of the particle number and its dependence on the choice of vacuum state.  
The different sets of annihilation and creation variables, \((a(\bm{p}), b^\dagger(-\bm{p}))\) and \((a(\bm{p}, t), b^\dagger(-\bm{p}, t))\), which are associated with the basis functions \((f(\bm{p}, t), f^*(\bm{p}, t))\) and \((\chi(\bm{p}, t), \psi(\bm{p}, t))\), are related through a canonical transformation known as the Bogoliubov transformation \(\mathcal{B}(t)\). The \(\bm{p}\)-component of this transformation, denoted \(\mathcal{B}(\bm{p}, t)\), is expressed as

\begin{align}
\begin{pmatrix}
a( \bm{p},t) \\
b^\dagger( -\bm{p},t)
\end{pmatrix}
&=
\mathcal{B}(\bm{p},t)
\begin{pmatrix}
a( \bm{p}) \\
b^\dagger( -\bm{p})
\end{pmatrix},
\quad 
\mathcal{B}(\bm{p},t) =\begin{pmatrix}
\alpha(\bm{p},t)& \beta^*(\bm{p},t) \\
\beta(\bm{p},t) & \alpha^*(\bm{p},t).
\label{eqn22}
\end{pmatrix}
\end{align}
The preservation of the commutation relations of the creation and annihilation variables ensures that the Bogoliubov coefficients satisfy
\begin{align}
     |\alpha(\bm{p},t)|^2 - |\beta(\bm{p},t)|^2 &= 1.
     \label{eqn23}
\end{align}

The Bogoliubov transformation \eqref{eqn22} offers a clearer understanding of the physical implications of choosing annihilation and creation operators, as noted in the literature \cite{Blaschke:2017igl,Blaschke:2005hs}.
In this context, the time-dependent particle number \(\mathcal{N}(\bm{p},t)\) is defined for each mode \(\bm{p}\) in the quantum theory, which is characterized by the set \((\hat{a}(\bm{p},t), \hat{b}^\dagger(-\bm{p},t))\). This particle number is measured with respect to the reference vacuum state \(\ket{0}\), and is expressed as:
\begin{align}
   \mathcal{N}(\bm{p},t) &= \langle 0|\hat{a}(\bm{p},t)^{\dagger}\hat{a}( \bm{p},t)|0\rangle =  \langle 0|\hat{b}^{\dagger} (-\bm{p},t) \hat{b}(-\bm{p},t)|0\rangle \\
   \nonumber
   &=|\beta( \bm{p},t)|^2.
\label{eqn24}
\end{align}
In the last step, we used the fact that the vacuum is annihilated by the original creation and annihilation operators, $ a (\bm{p}) \ket{0} = 0,$ and $b(-\bm{p}) \ket{0} =0,$ , under the assumption that no particles were initially present \cite{Kluger:1991ib}. The total number of particles created in mode \( \bm{p} \) is given by the final value of $\mathcal{N} (\bm{p},t) :$

\begin{equation}
     \mathcal{N}(\bm{p},t= + \infty ) = |\beta(\bm{p},t= + \infty )|^2.
     \label{Np}
\end{equation}
Consequently, $ \mathcal{N}(\bm{p},t)$ can be interpreted as the number of real particles only at asymptotic times, when the external field is switched off (or, vanished). Alternative interpretations that overlook this peculiarity of $ \mathcal{N}(\bm{p},t)$  can lead to surprisingly unconventional results \cite{Blaschke:2005hs,Gregori:2010uf}.
\newline
In this work, we focus on the complete time evolution of the particle number  $\mathcal{N}(\bm{p},t),$ tracing its progression from an initial value of zero to a final asymptotic value $\mathcal{N}(\bm{p},t= + \infty )$.
\par
To express the Bogoliubov coefficients, we use the classical equivalence between $\phi(\bm{p},t)$ and $\pi(\bm{p},t)$, expressed in terms of an exact solution $f(\bm{p},t)$ (equation \eqref{eqn11}) and in terms of \(\chi(\bm{p},t)\) and \(\psi(\bm{p},t)\) (equation \eqref{eqn13}). 
We finally deduce that
\begin{align}
    \mqty(\alpha(\bm{p},t)\\\beta^*(\bm{p},t)) &= 
    M^{-1} 
    \mqty(f(\bm{p},t)\\ \dot{f}(\bm{p},t)),
    \notag\\
    M^{-1}(t) &= 
    i\mqty(\psi^*(\bm{p},t) & -\chi(\bm{p},t)\\     
    -\psi(\bm{p},t) & \chi(\bm{p},t)).
\label{eqn25}
\end{align}
For the $\mathcal{N}(\bm{p},t)$ first we can write the bogoliubov  coefficient  $|\beta(\bm{p},t)|^2$ in terms of this particular mode function $(\chi(\bm{p},t), \psi(\bm{p},t))$   corresponding to the reference  vacuum (or,equivalent to the instantaneous adiabatic vacuum) \cite{Blaschke:2017igl,Dabrowski:2016tsx}.
\begin{align}
    |\beta(\bm{p},t)|^2 =    |\psi(\bm{p},t)|^2  |f(\bm{p},t)|^2 - \dot{f}(\bm{p},t)  [  \psi \chi^*(\bm{p},t) f(\bm{p},t) +  \psi^*(\bm{p},t) \chi(\bm{p},t) f^*(\bm{p},t) ] + |\chi(\bm{p},t)|^2 |\dot{f}(\bm{p},t)|^2. 
    \label{eqn26}
\end{align}
Also, one can now express $\mathcal{N} (\bm{p},t) $ in terms of the functions $\mathcal{W}(\bm{p},t)$,$\Theta(\bm{p},t)$, and  $\mathcal{Y} (\bm{p},t)$, which characterize \(\chi(\bm{p},t)\) and \(\psi(\bm{p},t)\), and the exact mode solution \(f(\bm{p},t)\) that defines the reference vacuum \(\ket{0}\):
\begin{align} 
    \mathcal{N}(\bm{p},t)&=\frac{\mathcal{W}(\bm{p},t)}{2}\left[1+\mathcal{Y}(\bm{p},t)^2\right]|f(\bm{p},t)|^2+\frac{1}{2\mathcal{W}(\bm{p},t)}|\dot{f}(\bm{p},t)|^2
    \notag\\
    &-\frac{1}{2}+\mathcal{Y}(\bm{p},t)\Re{f^*(\bm{p},t)\dot{f}(\bm{p},t)}.
\label{eqn28}
\end{align}
This result generalizes the expression found in 
\cite{Fedotov:2010ue,Ahmadiniaz:2022vqd}, which corresponds to the specific case where we choose  $( \chi(\bm{p},t), \psi(\bm{p},t))$ as the zeroth-order adiabatic approximation \((\chi^{(0)} (\bm{p},t), \psi^{(0)}(\bm{p},t))\), fixed by \eqref{eqn21} and \( \mathcal{Y}^{(0)} (\bm{p},t) = 0\):
\begin{equation} 
   \mathcal{N}^{(0)}(\bm{p},t)=\frac{\omega(\bm{p},t)}{2}|f(\bm{p},t)|^2+\frac{1}{2\omega(\bm{p},t)}|\dot{f}(\bm{p},t)|^2-\frac{1}{2}.
   \label{eqn29}
\end{equation}
In this case, \(f(\bm{p},t)\) is the zeroth-order adiabatic mode (with initial adiabatic conditions at \(t_0\)), and Eq.~\eqref{eqn29} represents the particle number interpretation that is equivalent to the quantum kinetic equation formalism approach.
In this case,$ f(\bm{p},t) $ represents the zeroth-order adiabatic mode, defined with initial adiabatic conditions at $ t_0$ . Equation \eqref{eqn29} provides an interpretation of the particle number that is equivalent to the quantum kinetic equation formalism approach \cite{Blaschke:2017igl,Kluger:1991ib}.
\subsection{Correlation function}
\label{corfun}
 To describe the creation and annihilation of particles in our physical system, we require the equation of motion for the recently derived time-dependent adiabatic particle number,  $\mathcal{N}(\bm{p},t)$. Consequently, it is necessary to define a term that captures these creation and annihilation processes.  
In this context, we discuss the time-dependent pair correlation function, which is given by
\begin{align}
     \mathcal{C}( \bm{p}, t) &=  \langle 0|\hat{a}( \bm{p},t)^{\dagger}\hat{b} ^\dagger(-\bm{p},t)|0\rangle   
     \label{eqn30}
     \end{align}
 and its complex conjugate
 \begin{align}
     \mathcal{C}^*( \bm{p}, t) &=  \langle 0|\hat{a}( \bm{p},t)\hat{b}(-\bm{p},t)|0\rangle   
     \label{eqn31}
\end{align}
As can easily be seen, this function $\mathcal{C}(\bm{p},t)$ consisting of creation operators for a particle and an anti-particle with the opposite momentum describes the process of creation  of pairs.
 Calculating the expectation value and doing the limiting procedure  one can find \cite{Blaschke:2017igl,Sah:2023jlz} 
 \begin{align}
        \mathcal{C}( \bm{p}, t) &=  \alpha(\bm{p},t) \beta^*(\bm{p},t)
        \label{eqn32}
 \end{align}
\par
We found  that the function $\mathcal{C}(\bm{p}, t)$ in the terms of the functions $\mathcal{W} (\bm{p},t)$ and  $\mathcal{Y}(\bm{p},t),$ which  characterize $\chi(\bm{p},t)$  and $\psi(\bm{p},t)$ , and exact mode function  solution $ f(\bm{p},t )$  
\begin{align}
      \mathcal{C}(\bm{p},t) &=  \frac{\mathcal{W}(\bm{p},t)}{2} \Bigl [  1 - \mathcal{Y}^2 (\bm{p},t)  + 2 \ii  \mathcal{Y} (\bm{p},t)   \Bigr] |f(\bm{p},t)|^2 - \frac{ 1}{2 \mathcal{W}(\bm{p},t)}  |\dot{f} (\bm{p},t)|^2
      \nonumber\\ &
      +  \frac{1}{2} ( \ii - \mathcal{Y} (\bm{p},t) )   \Bigl[ f(\bm{p},t) \dot{f}^* (\bm{p},t)  + f^*(\bm{p},t) \dot{f} (\bm{p},t)\Bigr]
      \label{eqn33a}
\end{align}


As we know, vacuum polarization effects play a crucial role in the process of pair production. This effect is described through the functions $u(\bm{p}, t) = \Re(\mathcal{C}(\bm{p}, t))$ and $v(\bm{p}, t) = \Im(\mathcal{C}(\bm{p}, t))$ \cite{Blaschke:2017igl,Banerjee:2018fbw}.
\section{Sauter-pulse electric field}
\label{sauter}

\begin{figure}
    \centering
    \includegraphics[width=0.4\linewidth]{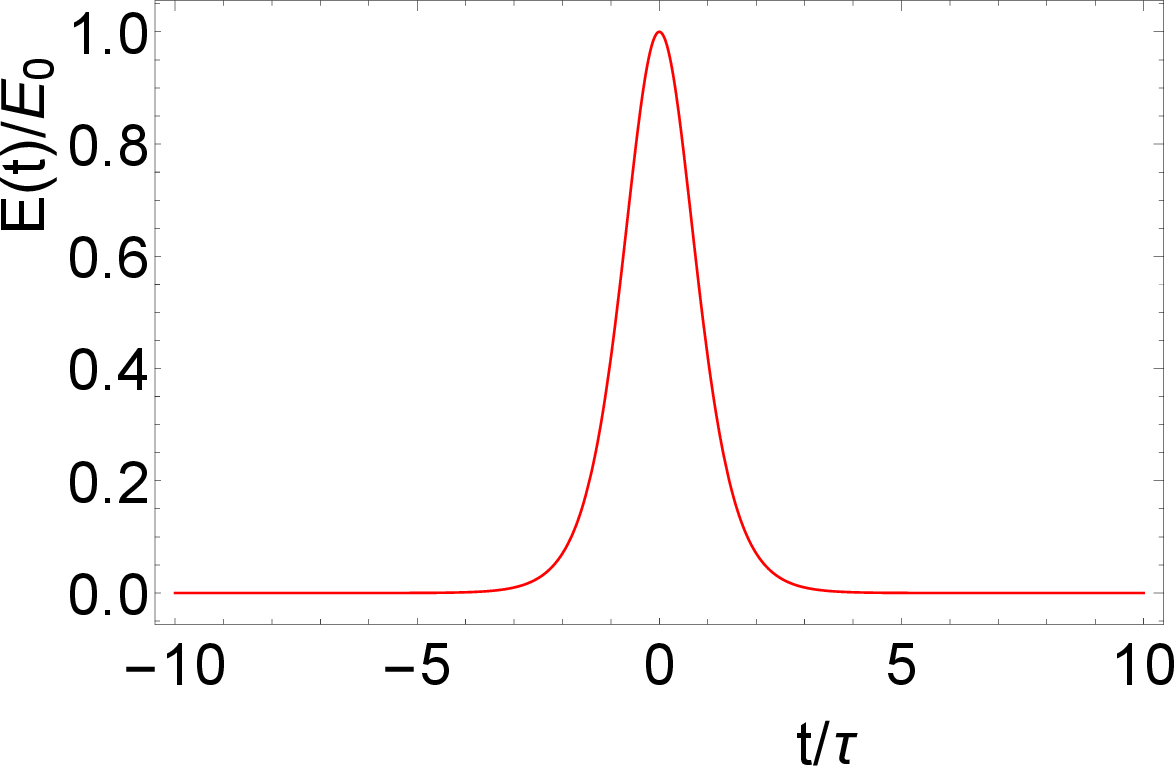}
    \includegraphics[width=0.4\linewidth]{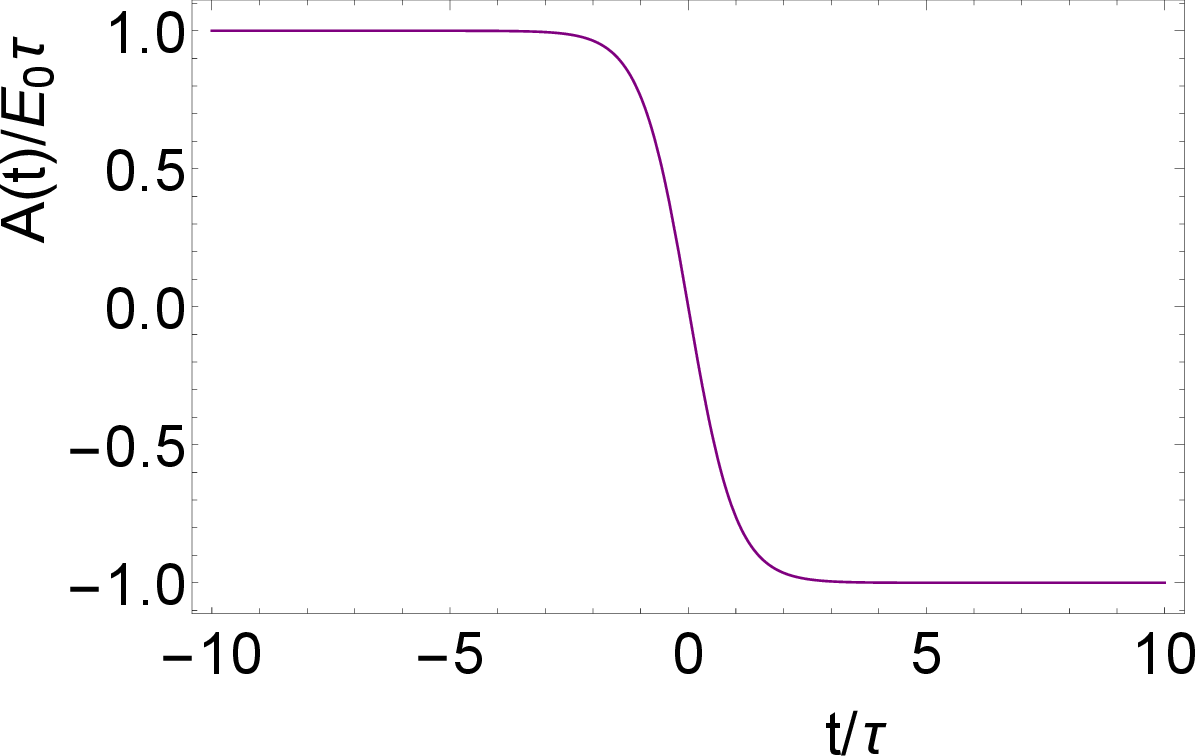}
    \caption{Temporal profile of the electric field \eqref{eqn33}  (left) and the corresponding vector potential with the choice $A(t=0) =0$(right) and all the units are taken in the
electron mass unit.}
    \label{fig:E_t_A_t}
\end{figure}
A spatially uniform external background is often used as an approximation for the electromagnetic field near the focal region of two counter-propagating laser pulses along the \(z\)-axis, forming a standing wave \cite{Blaschke:2008wf,Muller:2007df}. In general, pair production occurs near the peak of the electric field—comparable to the critical field limit—where the magnetic field vanishes. Although laser fields typically consist of many optical cycles, we consider a simplified model where the external field follows a Sauter profile, representing an extremely short laser pulse.

\begin{equation}
    E (t) =  E_0 \textrm{sech}^2 \left( \frac{t}{\tau} \right),
\label{eqn33}
\end{equation}

where \( \tau \) is the pulse duration and \( E_0 \) is the field strength. This electric field decays exponentially to zero for \( |t| \gg \tau \). In the limit \( \tau \to \infty \), the field becomes homogeneous in time. Choosing a gauge where \( A_0 = 0 \), the vector potential corresponding to this electric field is given by:
\begin{equation}
    A (t) = -E_0 \tau \tanh \left( \frac{t}{\tau} \right),
 \label{eqn34}
\end{equation}

The left panel of Fig.~\ref{fig:E_t_A_t} illustrates the temporal profile of the electric field in Eq.~\eqref{eqn33}.
\par
Sauter \cite{Sauter:1932gsa} originally analyzed the Dirac equation in the presence of an inhomogeneous scalar potential of the form \( V(z) = V_0 \, \text{sech}^2 (z/d) \). Since this problem reduces to a one-dimensional differential equation, a similar approach applies when the potential varies in time, as in Eq.~\eqref{eqn33}. The vacuum instability induced by a time-dependent Sauter-like electric field was first studied by Narozhny et al. in 1970 \cite{Narozhnyi:1970uv}. Since then, numerous works have revisited this topic, employing various theoretical approaches—including approximate methods—within this framework. Notable examples include Refs. \cite{Hebenstreit:2010vz,Gelis:2015kya} and references therein.
\newline
Now, Eq. \eqref{eom} can be rewritten in the presence of electric field Eq.\eqref{eqn33} 
\begin{equation}
\ddot{f}(\bm{p},t) +\omega^2(\bm{p},t) f( \bm{p},t)=0,
\label{eqn35}
\end{equation}
where,
 \begin{align}
     \omega(\bm{p},t)&= \sqrt{ \Biggl(p_\parallel- e E_{0}\tau \tanh \Bigl(\frac{t}{\tau} \Bigr) \Biggr)^{2}+p^{2}_{\perp}+ m^{2}}.
     \label{eqn36}
 \end{align}
 This equation can be solved by converting it into a hypergeometric differential equation\cite{abramowitz}, by changing the time variable to $y = \frac{1}{2} \left(1 +\tanh \Bigl(\frac{t}{\tau} \Bigr)\right).$ 
 \newline
 The new variable $y$ transforms the equation as
 \begin{align}
    \br{\frac{4}{\tau^2} y \br{1-y} \de{y} y \br{1-y} \de{y} + \omega^2(\bm{p},y) } f( \bm{p},y)  = 0.
\label{eqn37}
\end{align}
In this case, solutions can be written in terms of
hypergeometric functions \cite{abramowitz} :
\begin{align}
     f(\bm{p},y)  = \frac{1}{\sqrt{2 \omega_0}} y^{-i\tau\omega_0/2} (y-1)^{i\tau\omega_1/2}  \Hyper{a,b,c;y}
    \label{eqn38}
\end{align}
where, $\Hyper{a,b,c;y}$ is the hypergeometric function, and
\begin{eqnarray}
    a&=&\frac{1}{2}+\frac{i}{2}(\tau\omega_{1}-\tau\omega_{0})-i\lambda,\nonumber\\
 b&=&\frac{1}{2}+\frac{i}{2}(\tau\omega_{1}-\tau\omega_{0})+i\lambda,\nonumber\\
 c&=&1-i\tau \omega_{0},\nonumber\\
 \lambda&=&\sqrt{(e E_{0}\tau^{2})^{2}-\frac{1}{4}}, \nonumber\\ \omega_{0}&=&\sqrt{(p_{\parallel}+ e E_0\tau)^2+p^{2}_{\perp}+m^2},\nonumber\\
 \omega_{1}&=&\sqrt{(p_{\parallel} -e E_{0}\tau)^{2}+p^{2}_{\perp}+m^2}.
 \label{eqn39}
\end{eqnarray}

It is convenient to study the pair creation problem in two dimensions.  Throughout this work, we set the transverse momentum to zero, \( p_\perp = 0 \), and define \( p_\parallel = p \), restricting our analysis to (1+1)-dimensional Minkowski spacetime.  

From Eq.~\eqref{eqn29}, the time-dependent particle number for each mode \( p \) is given by
   \begin{align}
   \mathcal{N}(p,t) &=  \frac{1 }{ 4   \omega_0 \mathcal{W}(p,t)}  \bigg| \frac{2}{\tau} y (1-y)  \frac{a b}{c} g_1 + \ii ( \mathcal{W}(p,t) ( \ii  + \mathcal{Y}(p,t) )- (1-y) \omega_0 - y \omega_1 ) g_2 \bigg|^2 
     \label{eqn40a}
\end{align}
where,
$ g_1 = \Hyper{1+a,1+b,1+c;y}$, $g_2 =\Hyper{a,b,c;y}.$

\begin{figure}[t]
\begin{center}
{\includegraphics[width =3.25in]{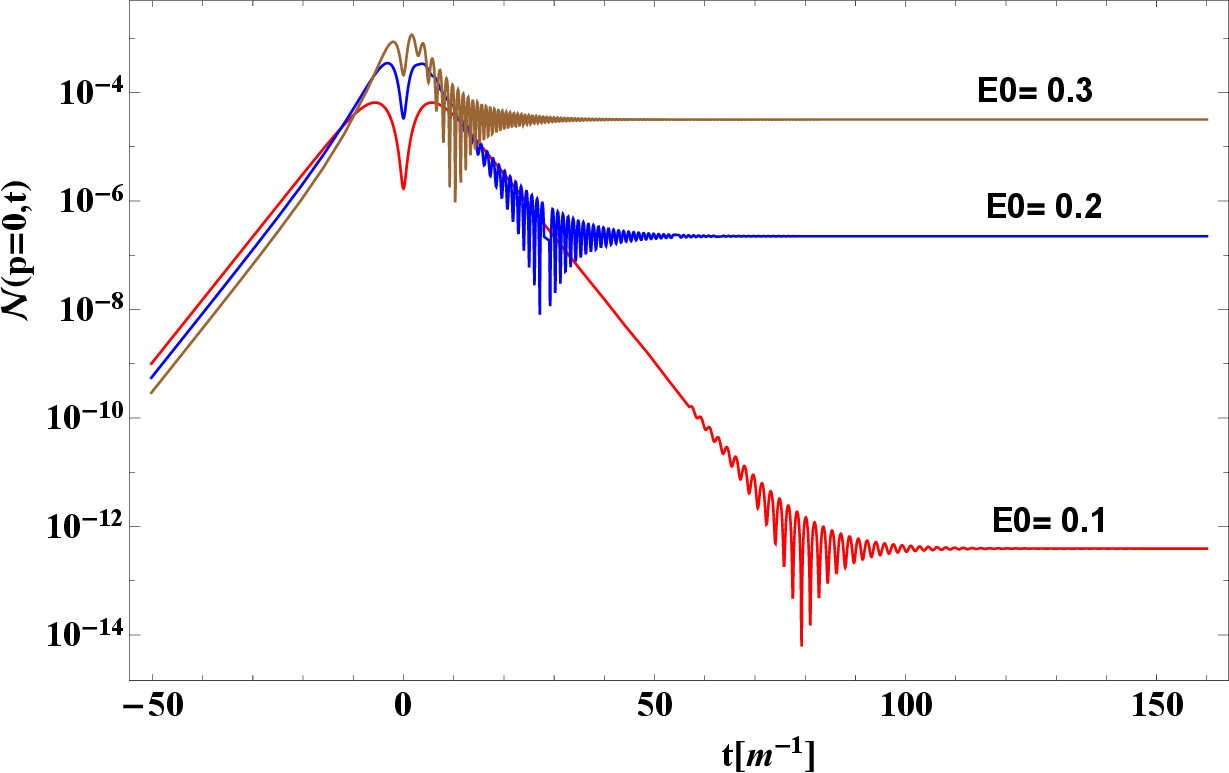}

}
\caption{Evolution of $\mathcal{N}(p,t)$ for different field strengths $E_0 = 0.1$ (red), $0.2$ (blue) ,$0.3$(brown) and  $ \tau =15 [m^{-1}].$ The momentum is considered to be zero, and all the units are taken in the electron mass unit.}
\label{N_t_E0}
\end{center}
\end{figure}
\begin{figure}[t]
\begin{center}
{
\includegraphics[width =3.25in]{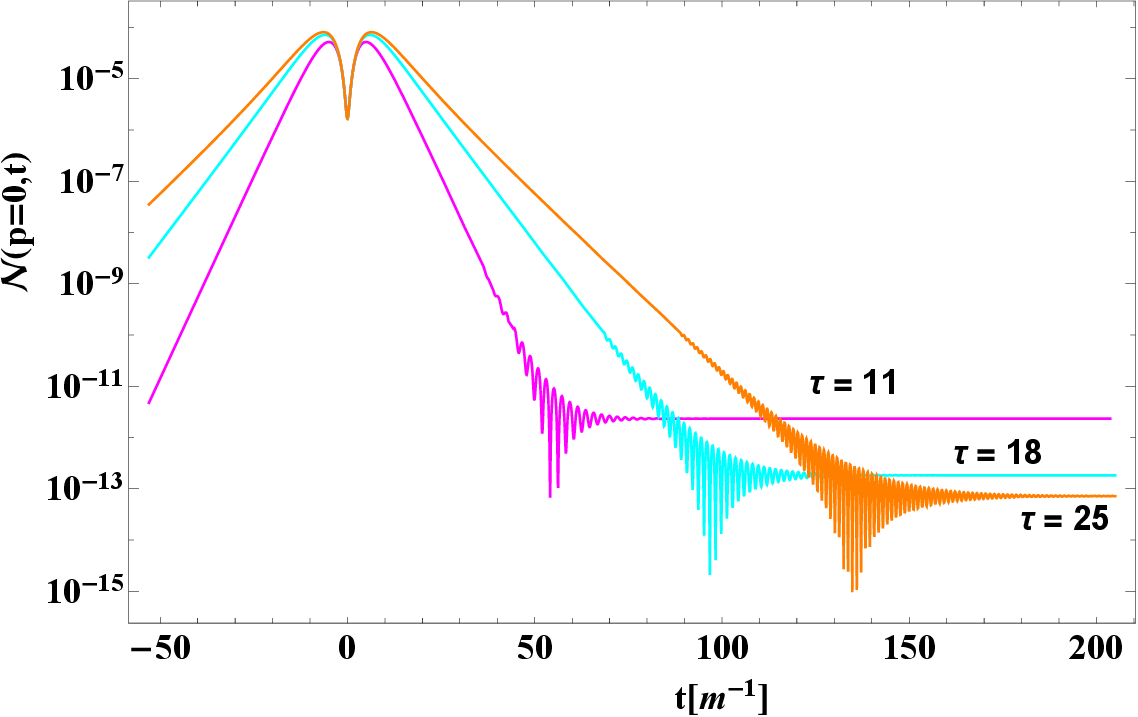}
}
\caption{Evolution of $\mathcal{N}(p,t)$ for different pulse duration $\tau = 11$ (magenta), $18$ (cyan) ,$25$(orange) and  $ E_0 =0.1E_c.$ The momentum is considered to be zero,and all the units are taken in the electron mass unit.}
\label{N_t_tau}
\end{center}
\end{figure}

\section{Numerical Results in $ (1+1)$ Dimensions}

We aim to analyze the dynamical features of the key quantity or functions responsible for pair production, as defined previously in section \ref{key qunatity}. To achieve this, we adopt the natural choice \( \mathcal{W}(p,t) = \omega(p,t) \) and \( \mathcal{Y}(p,t) = 0 \), which is frequently used in the literature for the quantum kinetic equation describing pair creation in intense laser fields (see reference \cite{Kluger:1991ib,Smolyansky:1997fc,Dumlu:2010ua}).
\subsection{Temporal Evolution}
In this subsection, we explore the dynamical behavior of $\mathcal{N}(p,t)$ for natural choice $\mathcal{W}(p,t) = \omega(p,t)$ and $\mathcal{Y}(p,t)=0$, frequently used in studies of pair creation from intense laser fields. Under this choice, the created particle number $\mathcal{N}(p,t)$ is given by
\begin{align}
   \mathcal{N}(p,t) &=  \frac{1 }{ 4   \omega_0 \omega(p,t)} 
\Biggl(\bigg|\frac{2}{\tau} y (1-y)  \frac{a b}{c} g_1 \bigg|^2 + \bigg| \bigl((\omega_0 + \ii \omega(p,t)) + y (\omega_1 - \omega_0)\bigr)g_2\bigg|^2 
\nonumber \\&+ 2 \Re{(\frac{2}{\tau} y (1-y)  \frac{a b}{c} g_1 \bigl((\omega_0 -\ii \omega(p,t)) + y (\omega_1 - \omega_0)\bigr)g^{*}_2)} \Biggr)
     \label{eqn40}
\end{align}
The figure \ref{N_t_E0} presents the time evolution of  particle number \( \mathcal{N}(p,t) \) for different electric field strengths \( E_0 = 0.1 \) (red), \( 0.2 \) (blue), and \( 0.3 \) (brown), with a pulse duration of \( \tau = 15 \, [m^{-1}] \). 
Initially, before the interaction with the external field (\( t < 0 \)), the particle number remains negligible, as expected in the absence of a background field. As the field is switched on, the particle number starts fluctuating, indicating the interaction of the quantum vacuum with the time-dependent electric field. The particle number increases monotonically, reaching its peak around $t \approx 0 $, where the electric field attains its maximum value. However, in the region \( t \le 0 \), \( \mathcal{N}(p,t) \) exhibits a valley and peak structure with a dip at \( t = 0 \), deviating from the direct behavior of the electric field. This contrasts with the fermionic (spin-1/2) case, where the evolution of \( \mathcal{N}(p,t) \) follows the electric field more closely (see ref. \cite{Blaschke:2014fca}).  

During the peak of the external pulse (\( t \approx 0 \)), the system exhibits strong oscillatory behavior, which corresponds to the continuous creation and annihilation of virtual pairs. The oscillations, known as Zitterbewegung, arise from quantum interference between dynamically evolving states, highlighting the complex nature of time-dependent pair creation.
The amplitude and frequency of these oscillations are directly influenced by the field strength, with higher values of \( E_0 \) leading to more pronounced variations in \( \mathcal{N}(p,t) \).  
As the field diminishes after the interaction period (\( t > 0 \)), different behaviors emerge based on the applied field strength. For \( E_0 = 0.1 \), the particle number decreases significantly and exhibits long-lived oscillations, suggesting that only a small fraction of created pairs remain as real particles while the rest undergo recombination. In contrast, for \( E_0 = 0.2 \), the particle number stabilizes at a finite value, indicating a saturation of the pair production process. For the highest field strength (\( E_0 = 0.3 \)), the final particle number is significantly larger, demonstrating the expected increase in pair production with stronger fields. This result aligns with the predictions of the Schwinger mechanism, where pair creation is exponentially suppressed for weaker fields but becomes more efficient as the field strength increases.  
\par
Figure \ref{N_t_tau} illustrates the time evolution of the particle number \( \mathcal{N}(p,t) \) for different pulse durations: \( \tau = 11 \) (magenta), \( \tau = 18 \) (cyan), and \( \tau = 25 \) (orange), with a fixed electric field strength of \( E_0 = 0.1E_c \). The plot provides insight into the creation and evolution of particle-antiparticle pairs under the influence of a time-dependent electric field.  A key feature observed in the Fig.\ref{N_t_tau} is the sharp increase in \( \mathcal{N}(p,t) \) around \( t = 0 \), where the electric field reaches its peak strength. The magnitude of this peak remains nearly identical across different pulse durations, indicating that the maximum value of \( \mathcal{N}(p,t) \) is primarily determined by the field strength at its peak (\( E(t \sim 0) \approx E_0 \)), regardless of how long the field is sustained.  
Another significant feature is the presence of oscillations in the particle number after the peak in the transient region. These oscillations are more pronounced for longer pulse durations (cyan and orange curves) compared to \( \tau = 11 \) (magenta). For shorter pulses, rapid variations in the electric field induce non-adiabatic transitions, resulting in stronger oscillatory behavior in \( \mathcal{N}(p,t) \). These oscillations gradually dampen as the field strength diminishes.  

A notable trend is that the final particle number, corresponding to \( \mathcal{N}(p,t) \) at large times (\( t > 100 \)), strongly depends on the pulse duration. Shorter pulses (magenta curve, \( \tau = 11 \)) yield a higher final number of pairs compared to longer pulses (cyan and orange curves). This suggests that prolonged interaction with the electric field reduces pair creation efficiency, as the system has more time for the produced pairs to evolve dynamically.  

For very short pulse durations (\( \tau = 11 \)), the oscillations decay more rapidly, and the final particle number stabilizes at a higher value than for longer pulses.  

Overall, the figure demonstrates the impact of pulse duration on pair production dynamics. While all cases exhibit a peak in particle number during the interaction, the extent of oscillations and the final yield vary significantly with the duration of the electric field. 
\begin{figure}[t]
\begin{center}
{
\includegraphics[width =3.5in]{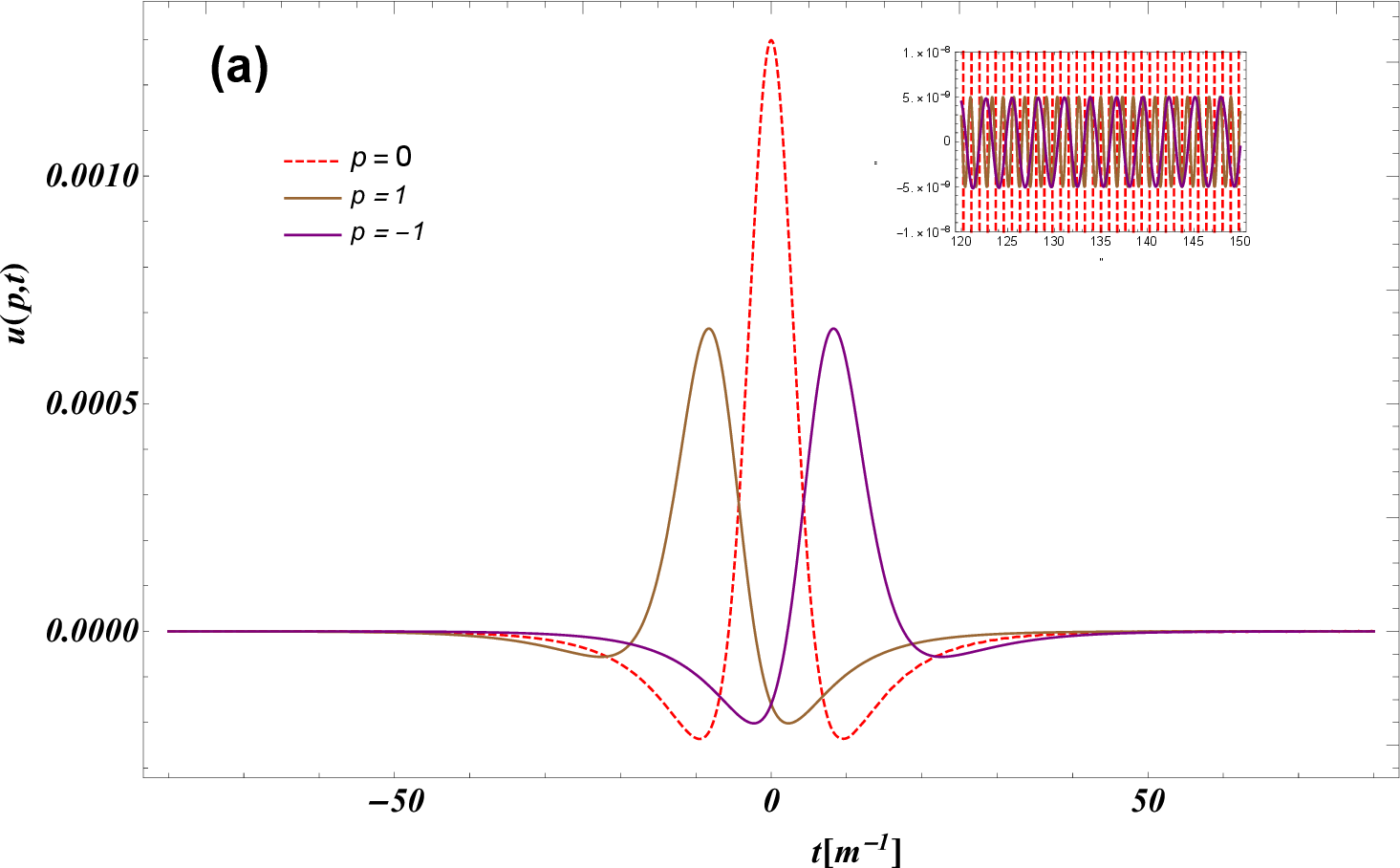}
\includegraphics[width =3.5in]{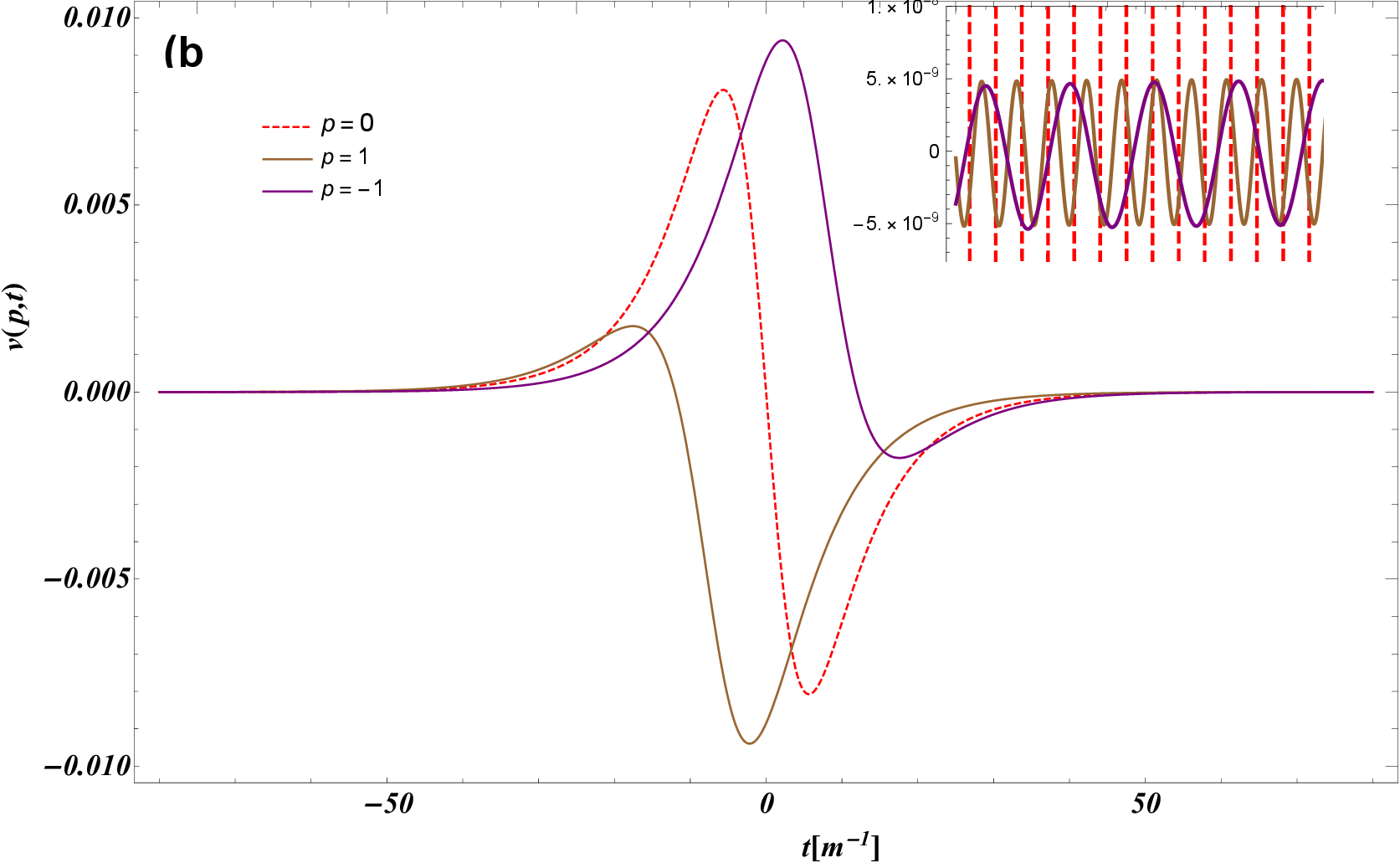}
}
\caption{\textbf{Left panel :}Evolution of  $u(t)$  and
\textbf{Right panel :}Evolution of  $v(t)$  for different momentum value $p$.The field parameters are  $E_0=0.1 E_c$ and $ \tau =15 [m^{-1}].$}
\label{u_v_p3}
\end{center}
\end{figure}
\begin{figure}[t]
\begin{center}
{
\includegraphics[width =3.5in]{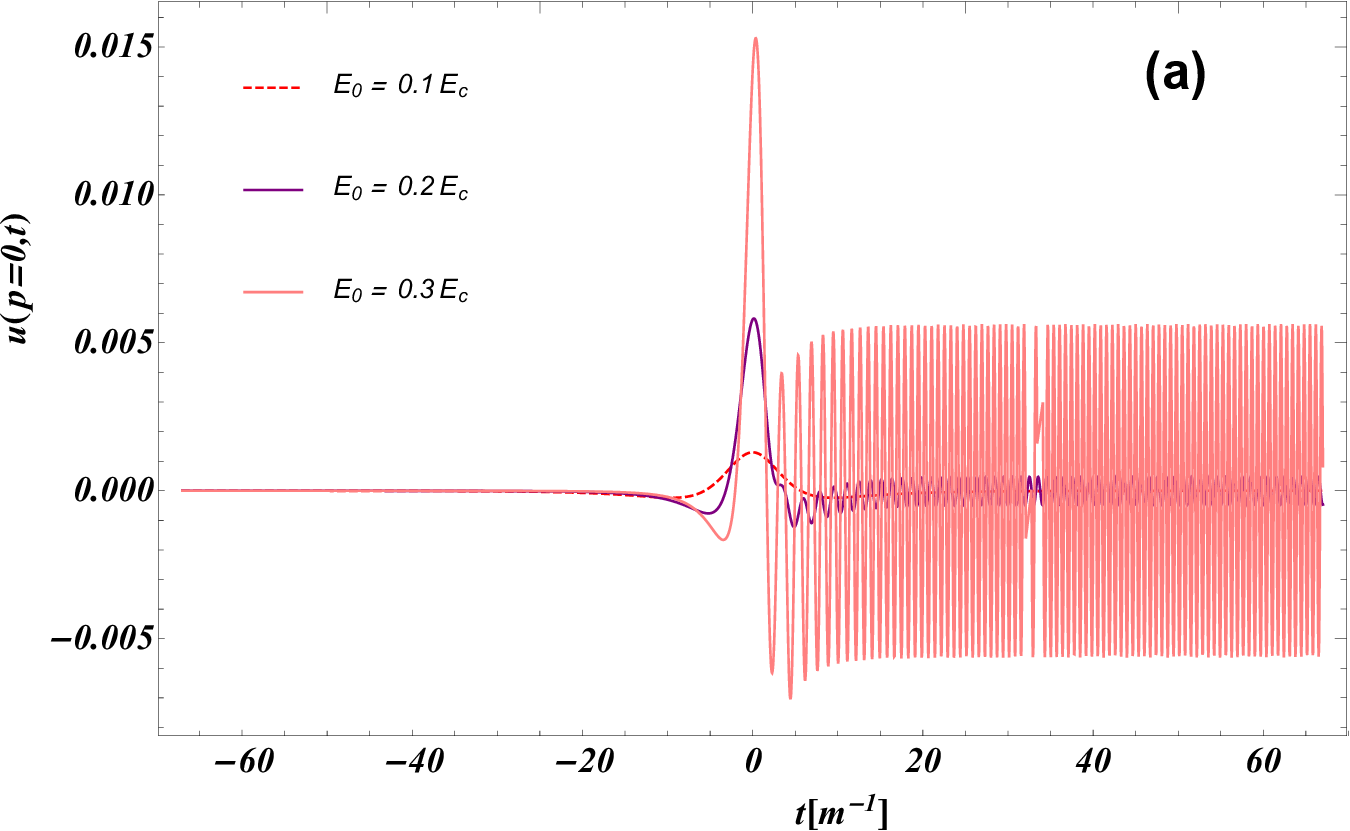}
\includegraphics[width =3.5in]{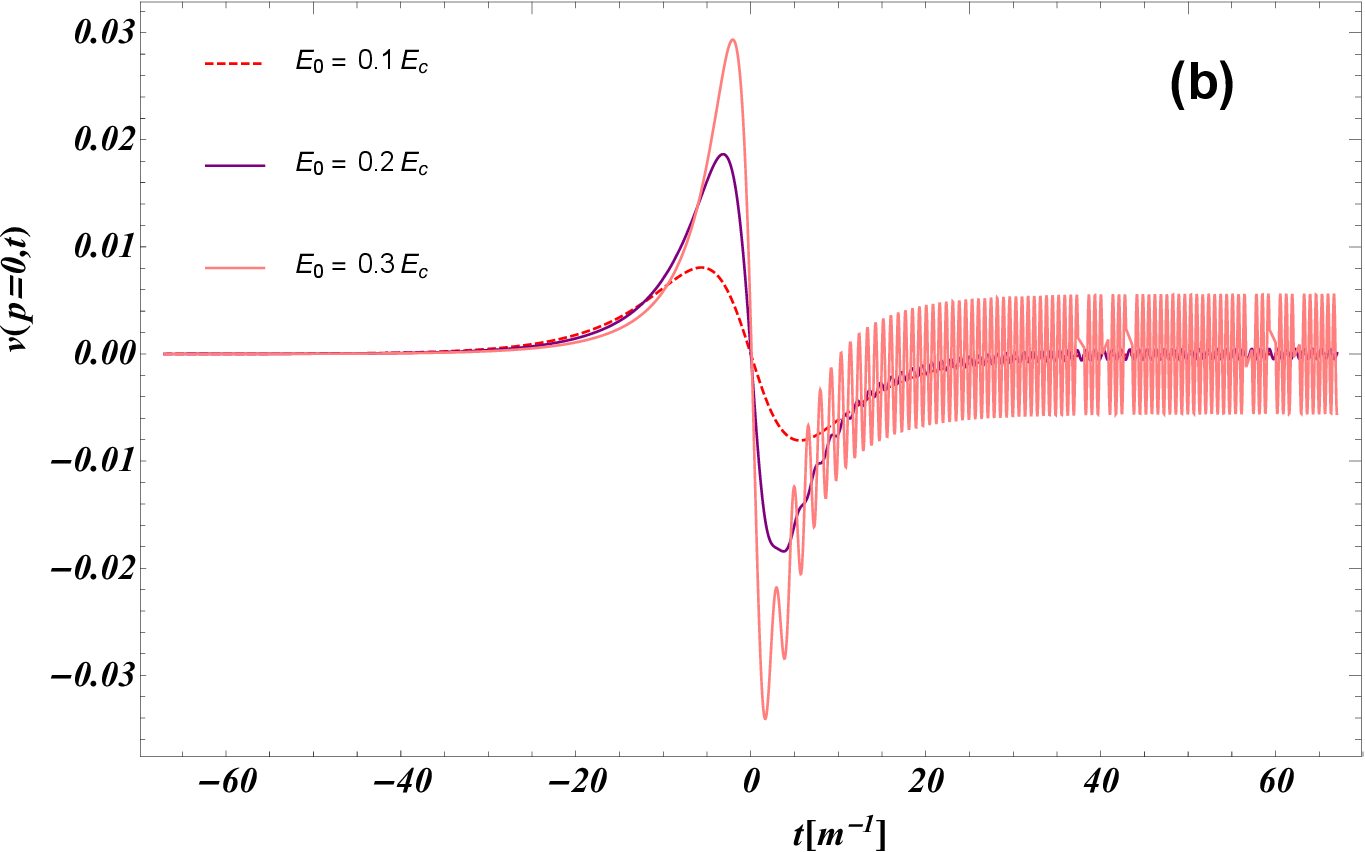}
}
\caption{\textbf{Left panel :}Evolution of  $u(t)$  and
\textbf{Right panel :}Evolution of  $v(t)$  for different field strengths $E_0$ .The
momentum is considered to be zero,and pulse duration, $\tau =15.$}
\label{u_v_E0}
\end{center}
\end{figure}

\subsection{Vacuum Polarization Function}  
As discussed in subsection \ref{corfun}, the vacuum polarization effects can be described by the functions $u(p,t)$ and $ v(p,t)$. The function \( u(p,t) \) can be interpreted as a current polarization function, while \( v(p,t) \) determines the energy density of the vacuum polarization 
\cite{Aleksandrov:2020mez,Smolyansky:2016gmp}. Both of these components originate from correlation functions as defined in the equation \eqref{eqn32}. 
\par
In the presence of intense Sauter pulsed  field \eqref{eqn33}, we compute the vacuum polarization function using the exact solution of the mode function \eqref{eqn41}, given by  
 \begin{align}
     u(p,t) &= \frac{2}{\tau \omega_0 }  |\frac{ a b }{c}|^2 | \frac{\tau }{2} \Bigl(  \frac{- \ii c}{ a b} \Bigr)  [ \omega_0  (1-y)  + \omega_1 y ] g_1  + y (1-y) g_2| ^2  -\frac{1}{ 2 \omega_0  } \mathcal{W} 
     (\mathcal{W}^2 -  \mathcal{Y}^2 )  
      \nonumber \\ &+ \mathcal{Y} \mathcal{W} \frac{2}{\tau \omega_0} y (1-y)  \Re{ \frac{ a b }{c}  g_1
     ^* g_2}
     \label{eqn41}
\end{align}
whereas the  corresponding counter-term
\begin{align}
      v(p,t) &= \frac{-1}{  \omega_0}  \mathcal{W}^2  \mathcal{Y} |g_1|^2 + y (1-y)  \frac{1}{\tau \omega_0 } \mathcal{W} \Re{ (\frac{a b}{c} g_1^* g_2)}
      \label{eqn42}
\end{align}

\begin{figure}[t]
\begin{center}
{
\includegraphics[width =3.5in]{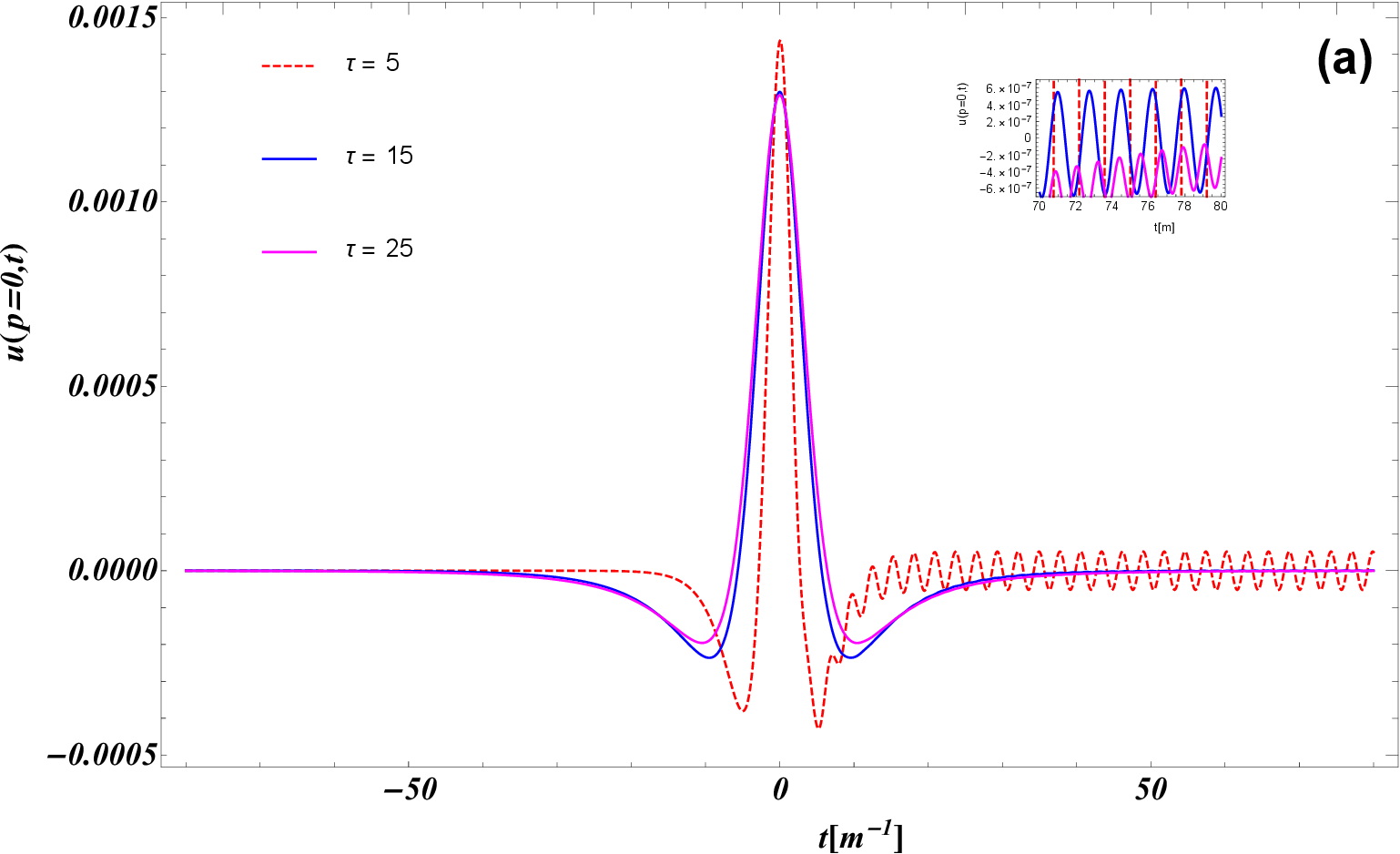}
\includegraphics[width =3.5in]{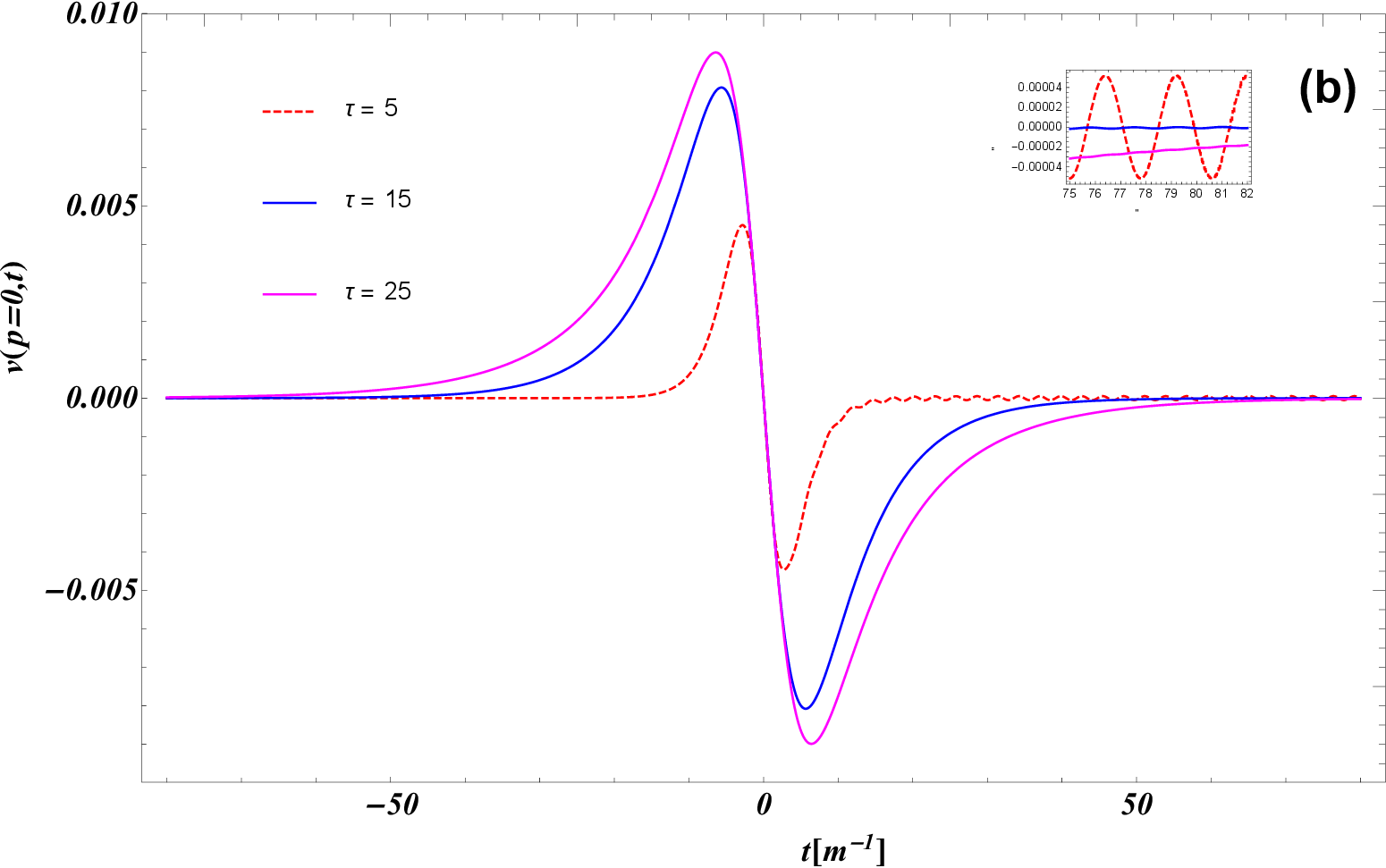}
}
\caption{\textbf{Left panel :}Evolution of  $u(t)$  and
\textbf{Right panel :}Evolution of  $v(t)$  for different pulse duration.
The momentum is considered to be zero, and field strength,  $E_0=0.1 E_c$.}
\label{u_v_tau}
\end{center}
\end{figure}

To analyze the dynamical behavior of the polarization functions $u(p,t)$ and $v(p,t)$,, we first fix the freedom in choosing $(\mathcal{W}(p,t),\mathcal{Y}(p,t))$. Following our previous approach, we adopt the natural choice $\mathcal{W}(p,t) = \omega(p,t)$ and $\mathcal{Y}(p,t)=0$.

Now, Equation \eqref{eqn41}-\eqref{eqn42} simplified as
 \begin{align}
     u(p,t) &= \frac{2}{\tau \omega_0 }  \bigg|\frac{ a b }{c} (y (\omega_1 -\omega_0)+ \omega_0) \bigg|^2 
     \Biggl[  \bigg|  \frac{y(1-y)}{(\omega_1 -\omega_0)y + \omega_0} g_2 \Bigg|^2 -\bigg| \frac{\tau c}{2 a b} g_1 \bigg|^2 - 2 \Re{\Bigl( \frac{y(1-y)}{y (\omega_1 -\omega_0)+ \omega_0}  \frac{\tau c}{2 a b}g_1 g^*_2 \Bigr)}  \Biggr]  
     \nonumber \\&
     -\frac{1}{ 2 \omega_0  } \omega(p,t)^3 |g_2|^2,
     \label{eqn41a}
\end{align}
and,
\begin{align}
      v(p,t) &=   y (1-y)  \frac{\omega(p,t)}{\tau \omega_0 }  \Re{( \frac{a b}{c} g_1^* g_2)}
      \label{eqn42a}
\end{align}


For a better understanding of the phenomenon of particle creation under a strong electric field, we will also trace the evolution of the vacuum polarization function $u(p,t),$ and its counterterm $v(p,t)$, which governs the depolarization.
The function \( u(p,t) \) remains nearly zero at early times (\( t < \tau \)), indicating that the system is initially in a stable vacuum state before the field interaction. A sharp peak appears at \( t = 0 \), with the strongest response occurring at \( p = 0 [m] \) (dashed red) for \( u(p,t) \)  and at \( p = -1 \) for \( v(p,t) \) (see Figure \ref{u_v_p3}). The structure of these peaks varies with momentum, reflecting the momentum dependence of the response. The insert of Fig.\ref{u_v_p3}(a) reveals residual oscillations that persist in late time where the field vanishes and the real particle-antiparticle is produced from vacuum.  The amplitude of oscillation is higher for lower momentum  $(p =0)$ value and oscillates about the origin. 
The function $v(p,t)$ exhibits an antisymmetric structure with respect to $t$ as shown in Fig. \ref{u_v_p3}(b). The function starts from zero at early times as similar to $u(p,t).$ The central peak for $p=0 $ (red dashed line) is symmetric about $t=0,$ but in an antisymmetric manner: its positive peak for $t >0 $ corresponds to a negative counterpart for $t < 0 $ 
For $p=1$ (brown) and $p=-1$ (purple), the structure remains antisymmetric, with one curve being a mirror image (with a sign flip) of the other.  As seen in the function $u(p,t)$ in asymptotic time, the oscillatory behavior is observed as seen in insert of Fig.\ref{u_v_p3}(b).
\par
We find out that in the early times, pair annihilation is stronger than pair creation, leading to a more prominent depolarization function $v(p,t)$ than the polarization function $u(p,t).$
This is evident from the larger amplitudes of $v(p,t)$ compared to $u(p,t)$ during the early evolution stage.
Both functions exhibit oscillations with varying amplitudes. The oscillations are particularly strong for $p=0,$ as shown in the insets of the plots.  The polarization function $u(p,t)$ develops a peak-like structure. This structure could be linked to the interference between vacuum fluctuations and field-induced pair creation. Unlike $u(p,t)$ the depolarization function $v(p,t)$ exhibits an antisymmetric peak structure.  In the transient stage, irregular oscillations are observed, likely due to the non-equilibrium dynamics of pair production and annihilation.
As time progresses, the system stabilizes, and the oscillations become regular and centered around zero.
At later times, $u(p,t)$ and $v(p,t) $ show similar oscillatory patterns, suggesting a balancing behavior between polarization and depolarization. This balance may indicate that the system is reaching a final state where creation and annihilation effects are in equilibrium as a result real particles are formed. This behavior of $u(p,t)$
and $v(p,t)$  closely resembles that observed for a fermion spin-\(\frac{1}{2}\) particle (see \cite{Banerjee:2018fbw,Sah:2023jlz}).
Next, we discuss the effect of the field strength, $E_0$ over the polarization function, $u(p,t)$ and $v(p,t).$ 
Figure \ref{u_v_E0} illustrates the temporal evolution of $u(p,t)$ and $v(p,t)$ for different field strengths. As the field strength $E_0$ increases, the peak magnitude of both $u(p,t)$ and $v(p,t)$ near $ t \approx 0$ becomes larger, indicating a stronger transient response. The oscillatory behavior emerge earlier and becomes more pronounced for $E_0 = 0.3 E_c,$ indicating the stronger dynamical evolution. The late-time behavior reveals oscillations, whose amplitude and duration increase with field strength.
The behavior of the polarization function $ u(p,t)$  and its counterterm  $ v(p,t)$ is also influenced by the pulse duration \( \tau \). An interesting observation is that at \( t \approx 0 \), the maximum peak value of \( u(p,t) \) occurs for the smaller pulse duration \( \tau = 5 \), whereas for \( v(p,t) \), the peak is highest for \( \tau = 25 \) (see Fig. \ref{u_v_tau}(a)-(b)). However, as the system approaches the residual final state, both \( u(p,t) \) and \( v(p,t) \) exhibit oscillations, with the amplitude being smaller for \( \tau = 5 \) compared to \( \tau = 25 \).
\begin{figure}[t]
\begin{center}
{\includegraphics[width =1.6519358802in]{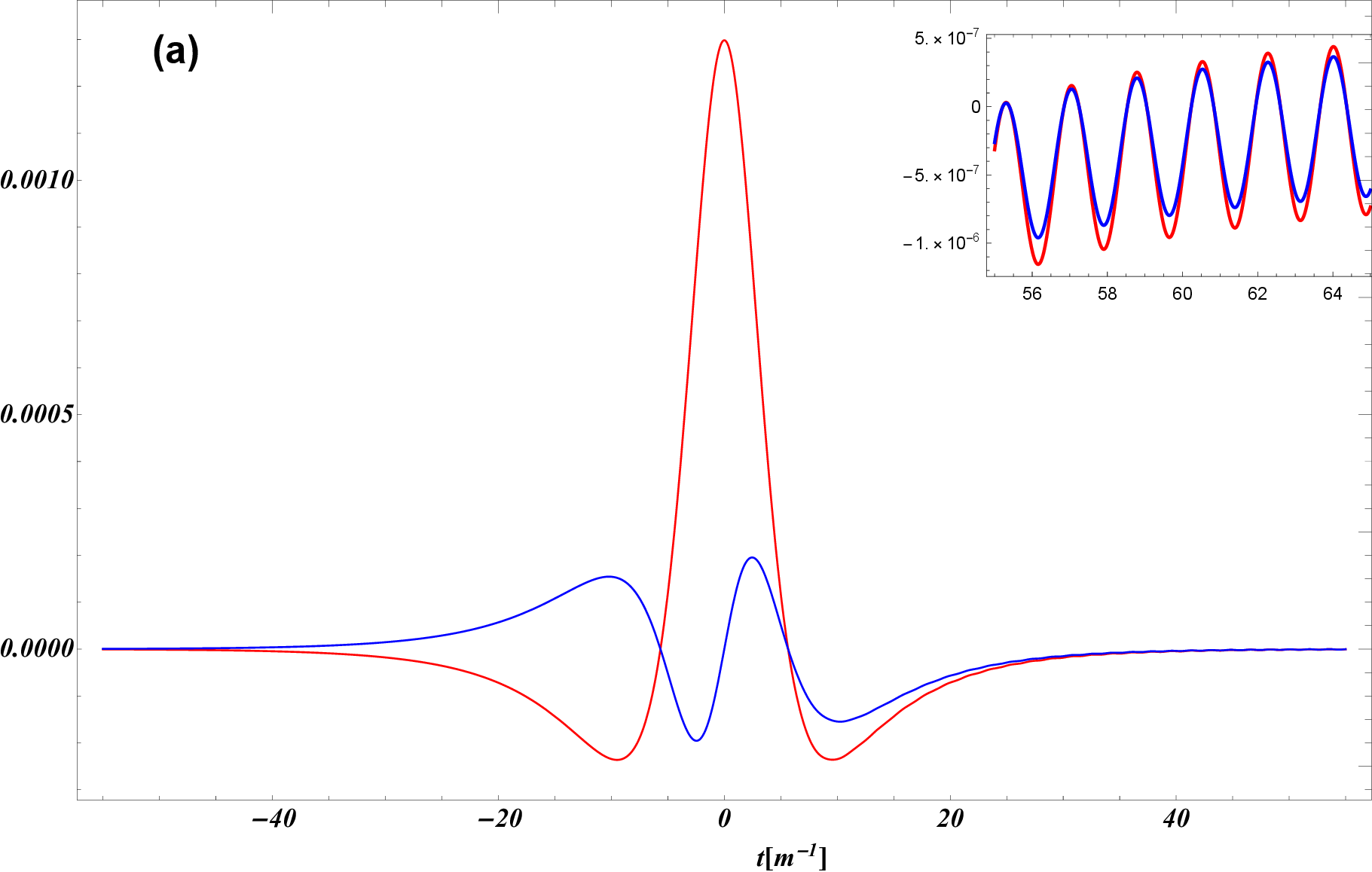}
\includegraphics[width =1.6519358802in]{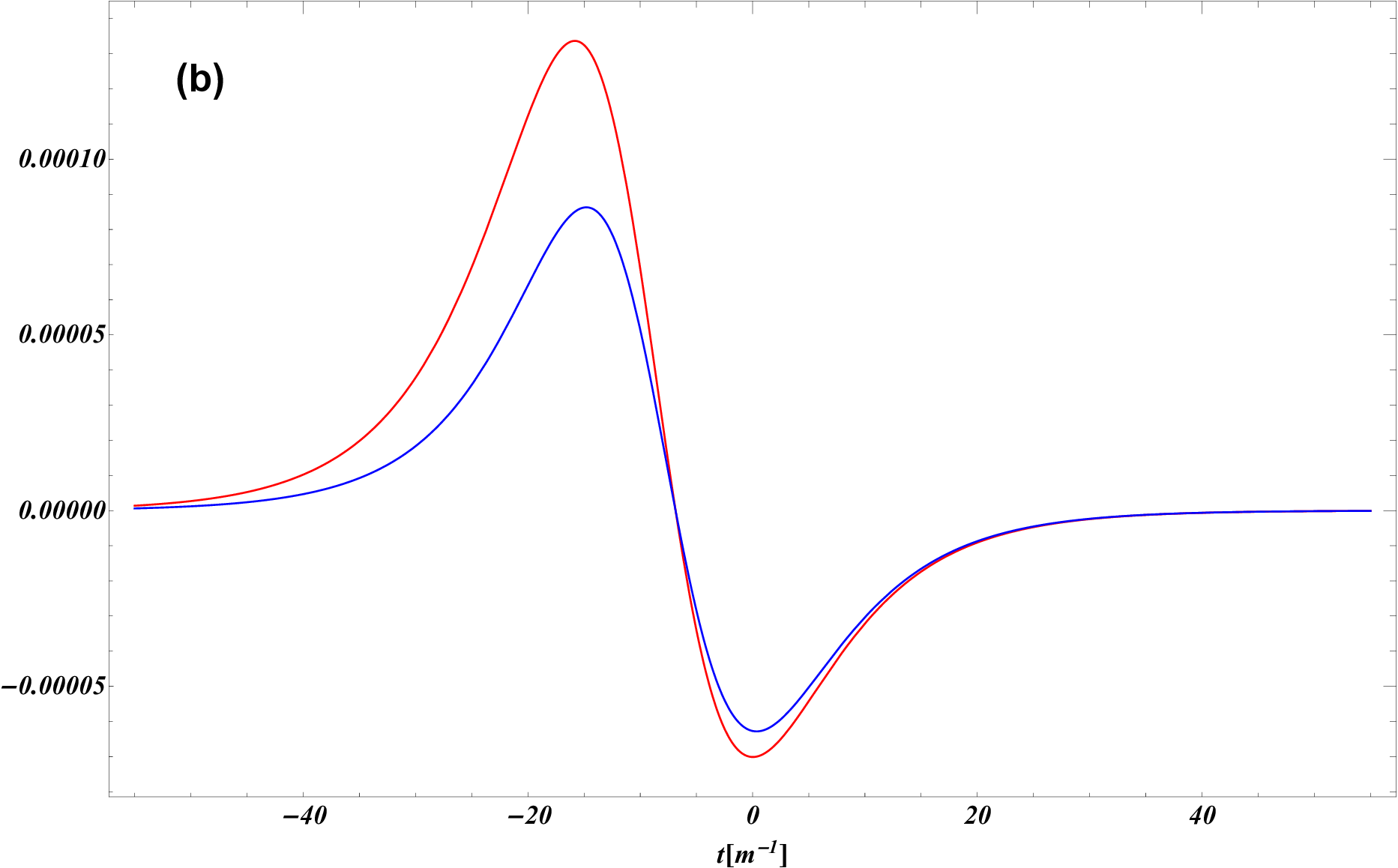}
\includegraphics[width =1.6519358802in]{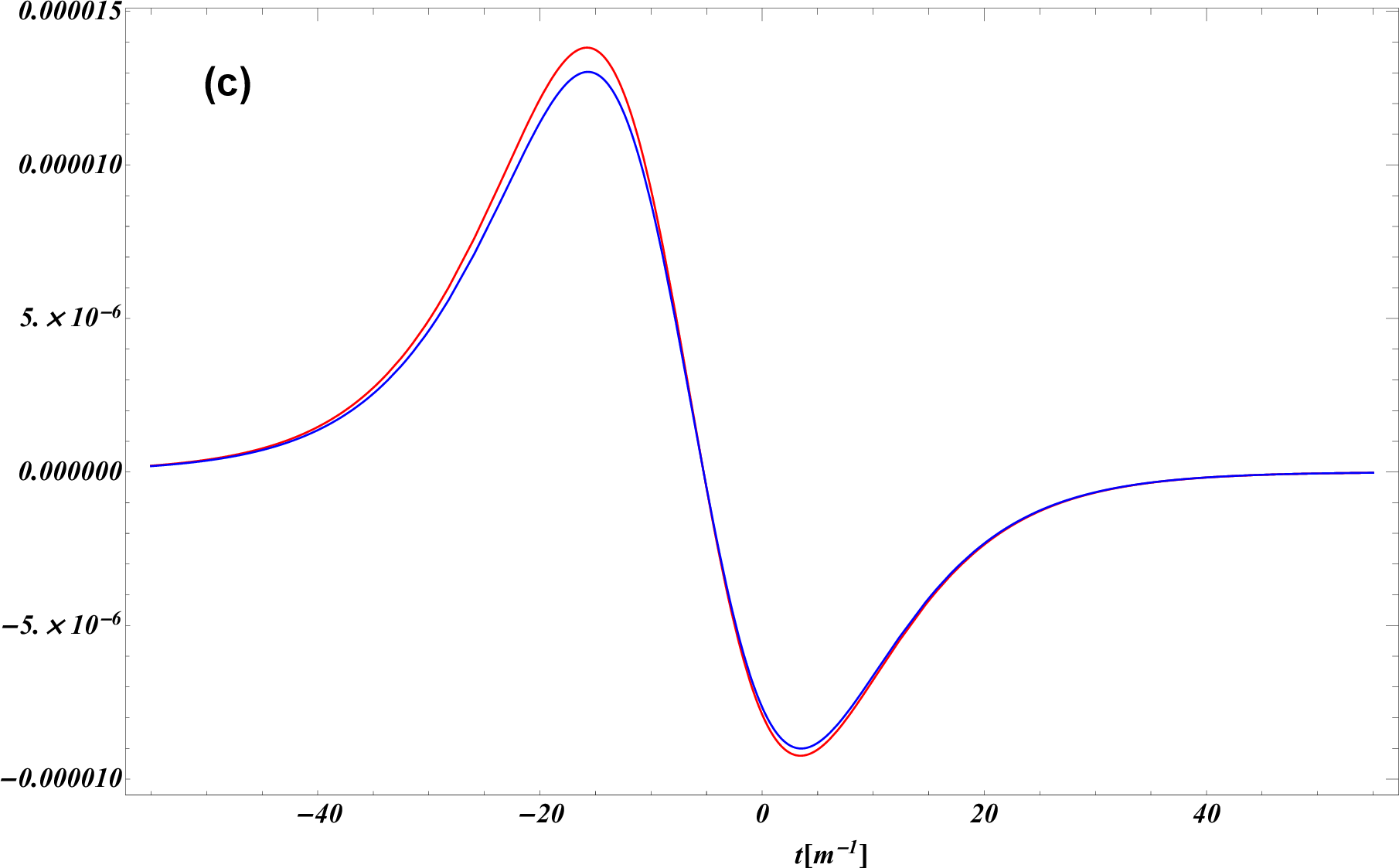}}
\caption{ The integrated of  $ J_{pol}(t)$  in \eqref{JP1} with blue curve and,  $u(p,t)$ (red curve) with different momentum values, (a) p =0, (b) p =2, 
(c) p=4.The field parameters are  $E_0=0.1 E_c$ and $ \tau =15 [m^{-1}].$ }
\label{J_p3_tau15}
\end{center}
\end{figure} 
\begin{figure}[t]
\begin{center}
{\includegraphics[width =1.6519358802in]{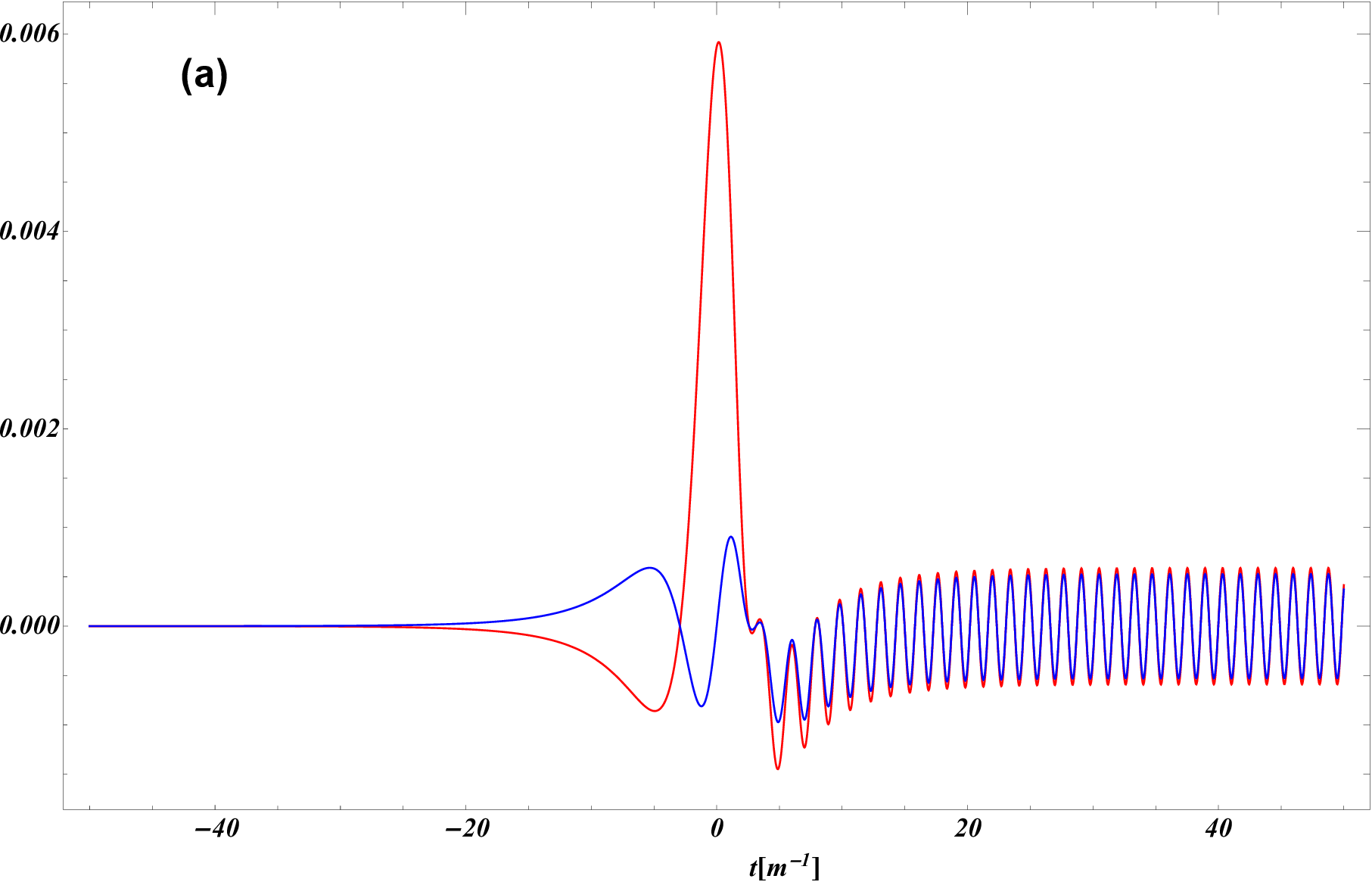}
\includegraphics[width =1.6519358802in]{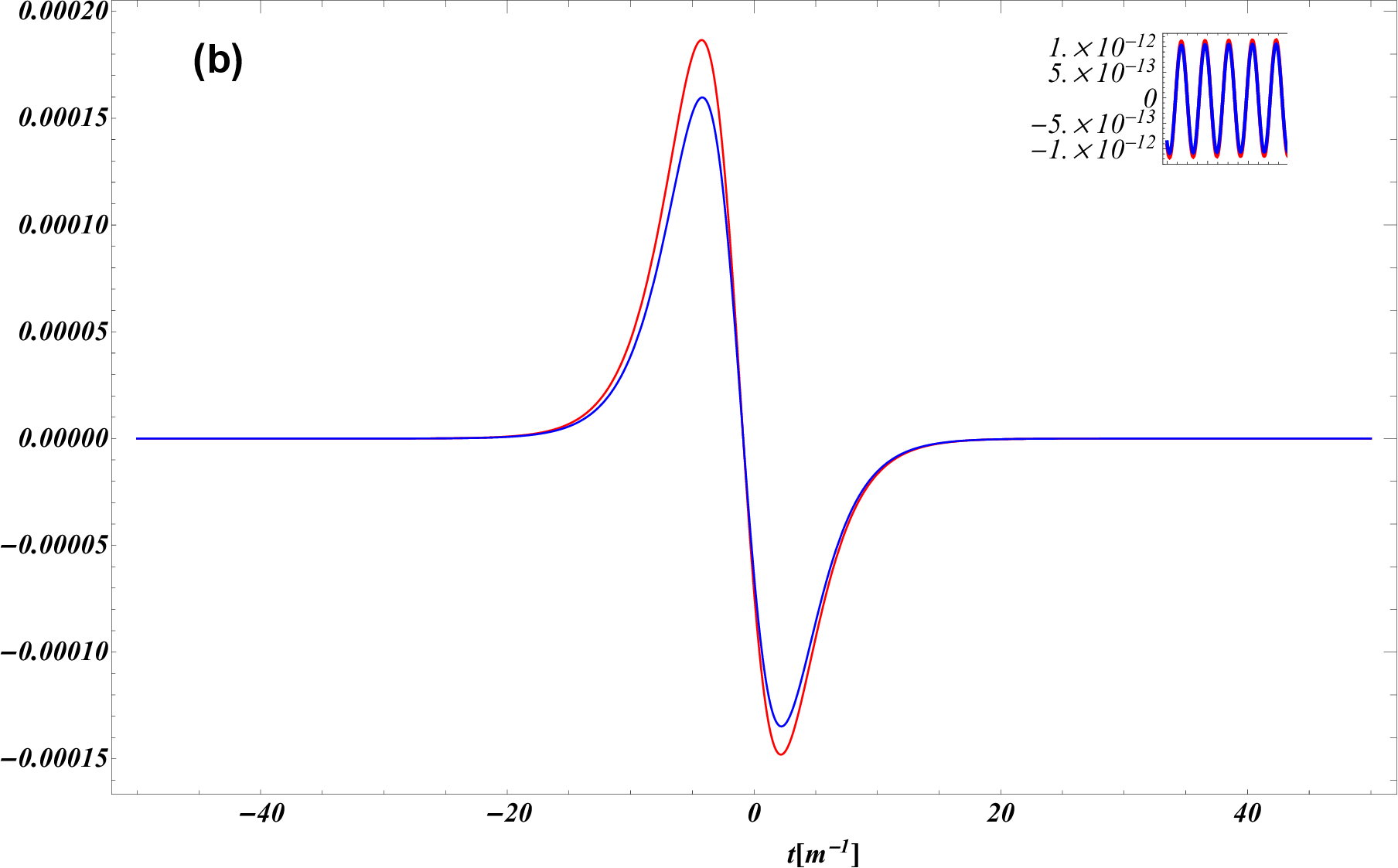}
\includegraphics[width =1.6519358802in]{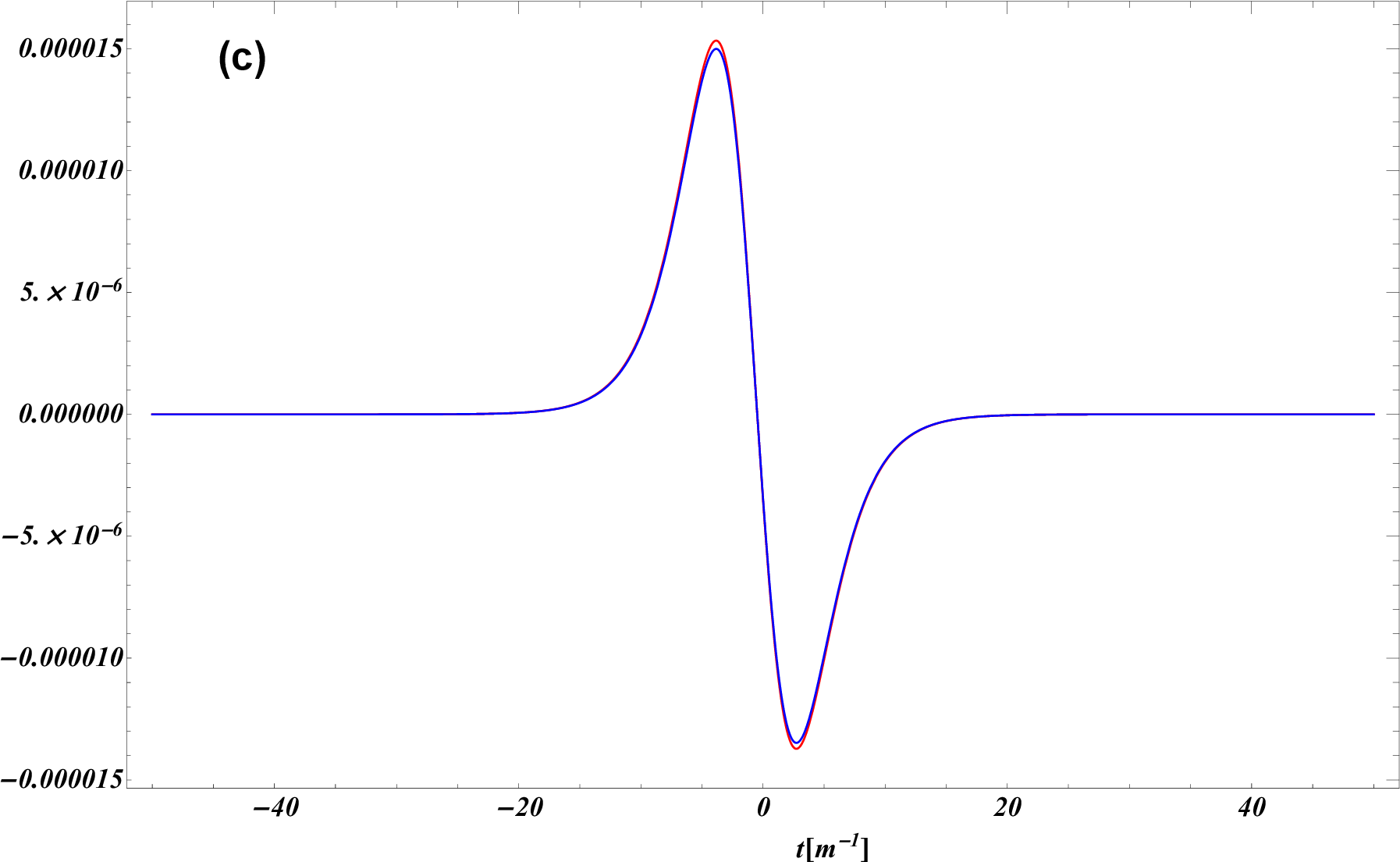}
}
\caption{ The integrated of  $ J_{pol}(t)$  in \eqref{JP1} with blue curve and,  $u(p,t)$ (red curve) with different momentum values, (a) p =0, (b) p =2, 
(c) p=4.The field parameters are  $E_0=0.1 E_c$ and $ \tau =5 [m^{-1}].$ }
\label{J_p3_tau5}
\end{center}
\end{figure} 
\subsection{Vacuum Polarization Current}
Next, we explore the dynamical properties of the vacuum polarization current, which depends on the polarization function \( u(p,t) \). The polarization current can be expressed as  
 \begin{align}
     J_{pol.}(t) &=   \frac{e}{2 \pi}\int dp \frac{(p - eA(t))}{\omega(p,t)} u(p,t).
    \label{JP1}
\end{align}
which originates from the classical equation $ J_{pol.}(t)= \frac{\partial}{\partial t} \mathcal{P}$, where$ \mathcal{P}$ is the Polarization field density, and hence is related to time derivatives of the number of created particles $\mathcal{N}(p,t)$ as discussed in the introduction section.

\par
We first examine the integrand of $ J_{pol.}(t)$ (see Eq.\eqref{JP1}) before performing the integration to gain a better understanding of the behavior of the polarization current. Through this one can see the relationship between the polarization response and momentum.
\begin{figure}[t]
\begin{center}
{\includegraphics[width =5.0in]{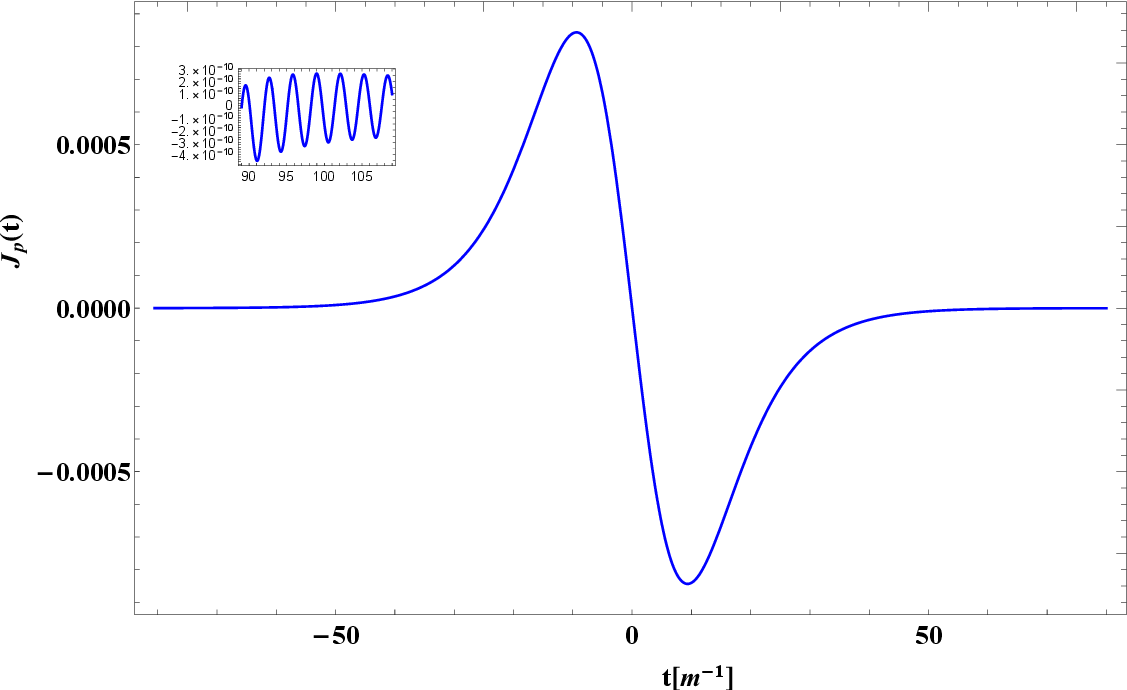}}
\caption{ Temporal dependence of polarization current  $ J_{pol}(t)$   defined  in \eqref{JP1}  for field parameter $E_0 = 0.2$ with $\tau =10[m^{-1}]$ .}
\label{J_t}
\end{center}
\end{figure} 
\begin{figure}[t]
\begin{center}
{\includegraphics[width =5.6519358802in]{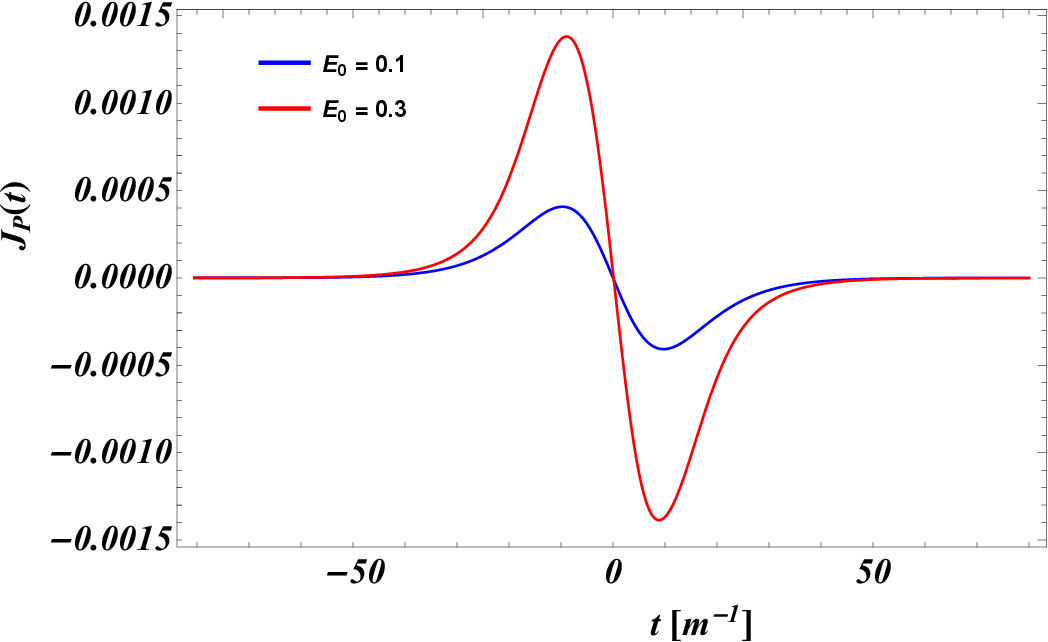}

}
\caption{ Temporal dependence of polarization current  $ J_{pol}(t)$   defined  in \eqref{JP1} for different field strength $E_0 = 0.1 $(blue),$0.3$(red) with $\tau =15[m^{-1}]$ .}
\label{J_t_E0}
\end{center}
\end{figure} 
Figure \ref{J_p3_tau15} illustrates the time evolution of the integrand of the polarization current $J_{pol.} (t)$ (blue) and the polarization function $u(p,t)$ (red) for different momentum values $p.$ The analysis is conducted for the field parameters $E_0 =0.1 E_c$ and $\tau = 15 [m^{-1}].$  Before the field becomes significant, both the integrand of polarization current and  $u(p,t)$ are close to zero for  $t << - 2 \tau.$ This indicates that there is no significant particle excitation in the system yet.
The integrand of $ J_{pol.}(t)$ and $u(p,t)$ both exhibit a well-defined peak structure, with the strongest response around $t = 0 [m^{-1}],$ corresponding to the peak of the field. The polarization current integrand shows a distinct oscillatory structure around the central peak. The function $u(p,t)$  follows a similar trend but exhibits a sharper peak. The blue curve, which involves the momentum-dependent prefactor, follows a similar trend of $u(p,t)$   but is modulated by $ \frac{(p- e A(t))}{\omega(p,t)}.$  
In the later time, $t > \tau$ the electric field vanishes, and $A(t)$ asymptotically approaches a constant value.
The integrand of $ J_{pol.}(t)$ in this time exactly follows the behavior of  $u(p,t)$ which exhibits regular oscillations about the origin. 
We also point out that as the momentum value increases ($p =2,4$ in the middle and right panels), the amplitude of both curves decreases, and the peak position shifts slightly, reflecting a momentum-dependent delay in the response.
The time evolution of the integrand of \(  J_{pol.}(t)\) and \( u(p,t) \) for \( p = 4 \) (see right panel) shows that both curves nearly overlap, indicating that at higher momentum, the integrand of \( J_P(t) \) behaves similarly to the function \( u(p,t) \). This can be understood as, for higher momentum values (\( p \gg e A(t) \)), the prefactor becomes negligible, and the integrand is dominated by \( u(p,t) \).
\newline
To demonstrate the dependence on the pulse duration $\tau$ we plot for $\tau =5 $ and compare with Fig.\ref{J_p3_tau15}  in the Fig.\ref{J_p3_tau5}.
For the shorter pulse duration, the integrand of $ J_{pol.}(t)$ (blue curve) exhibits more pronounced oscillations at $t > 0$, particularly for smaller momenta. This is because a shorter pulse contains a broader frequency spectrum.
The peak values are reduced compared to $\tau =15[m^{-1}],$ indicating that the system has less time to develop a strong response. The functions decay more quickly after the field pulse, meaning the system returns to equilibrium faster.
As seen in the right panel of Fig. \ref{J_p3_tau5}, for $p=4,$ the blue and red curves nearly overlap, similar to the $\tau =15 [m^{-1}]$ case. This indicates that at higher momentum, the behavior of the integrand of $J_P(t)$  remains dominated by $u(p,t)$ regardless of pulse duration. The regular oscillations at asymptotic times, after the field ceases, are observed with an increased frequency compared to the previous pulse duration.
\par
Now, to calculate the integration for obtaining the polarization current, we consider a one-dimensional case, integrating only over \( p = p_\parallel \), which lies along the direction of the electric field \eqref{eqn33}. The integration limits are chosen appropriately to ensure that the main contribution comes from the relevant momentum range, denoted as \( p_{\text{cut}} \). 
Figure \ref{J_t} shows the polarization current $J_P(t) = (\frac{2 \pi}{e}) J_{pol.} (t) $ computed by numerical integration, plotted as a function of time for field parameters, $E_0 = 0.2 E_c$ and $\tau = 10 [m^{-1}].$
The polarization current starts from zero and begins to increase as the electric field becomes significant. This is because the field induces virtual pair fluctuations, leading to a transient response.  The field reaches its maximum near $t =0,$ where it has the strongest influence on the system. The polarization  current exhibits a peak
(positive) followed by a dip(negative), corresponding to the response of the vacuum polarization under the influence of the strong field.
The sign change occurs due to the reversal in the field strength as $E(t)$ transitions from increasing to decreasing.
After the field diminishes, the polarization current does not vanish but exhibits regular oscillation similar to the trends shown by the  $u(p,t)$  in the late time region. This similar behavior is shown by the polarization pressure  $P_{pol.}$see the ref. \cite{Filatov:2007ha} for time-dependent mass configuration.  At asymptotic time the polarization current looks like undamped oscillation for $t >> \tau$  (see insert Fig. \ref{J_t}).

\begin{figure}[t]
\begin{center}
{\includegraphics[width =5.6519358802in]{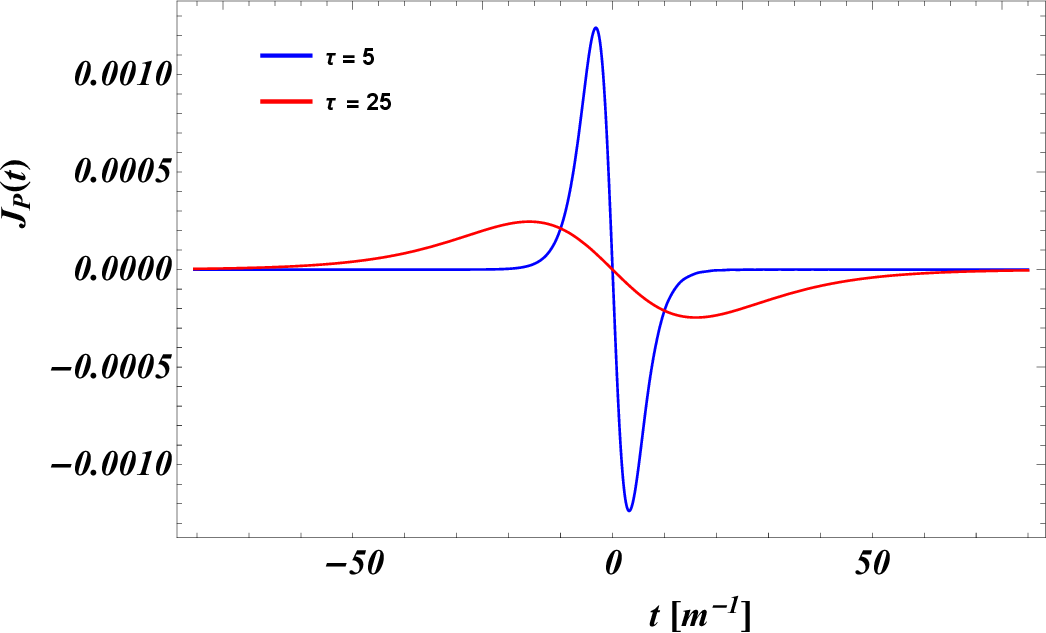}}
\caption{ Temporal dependence of polarization current  $ J_{pol}(t)$   defined  in \eqref{JP1} for different field strength $\tau = 5 $(blue),$25$(red) with $E_0 =0.1 E_c$ }
\label{J_t_tau}
\end{center}
\end{figure}

In the plot \ref{J_t_E0} shows the temporal evolution of the polarization current $J_P(t)$ for different field strengths. The red curve $(E_0 = 0.3 E_c)$ has a significantly larger peak and overall magnitude compared to the blue curve $(E_0 = 0.1 E_c).$ This is expected because the polarization current is driven by the external field, and a strong field induces a larger current response.
The overall shape of both curves remains similar, indicating that the polarization current follows a similar evolution regardless of field strength.
However, the stronger field $(E_0 = 0.3 E_c)$ leads to a sharper rise and decay, implying a more intense response in particle dynamics.  
Figure \ref{J_t_tau} represents the polarization current $J_P(t)$  for different pulse durations. For a shorter pulse ( $\tau=5,$ blue curve ), the current is sharply peaked and localized in time, with a higher peak amplitude due to the strong response to the rapidly varying field. 
In contrast, for a longer pulse ($\tau = 25$, red curve ), the response is smoother and more extended in time. The peak amplitude is lower, as the field varies more gradually. 
At late times ( $ t >> \tau $), the current gradually decays but residual oscillation persists, particularly for the shorter pulse.



\begin{figure}
    \centering
    \includegraphics[width=0.395\linewidth]{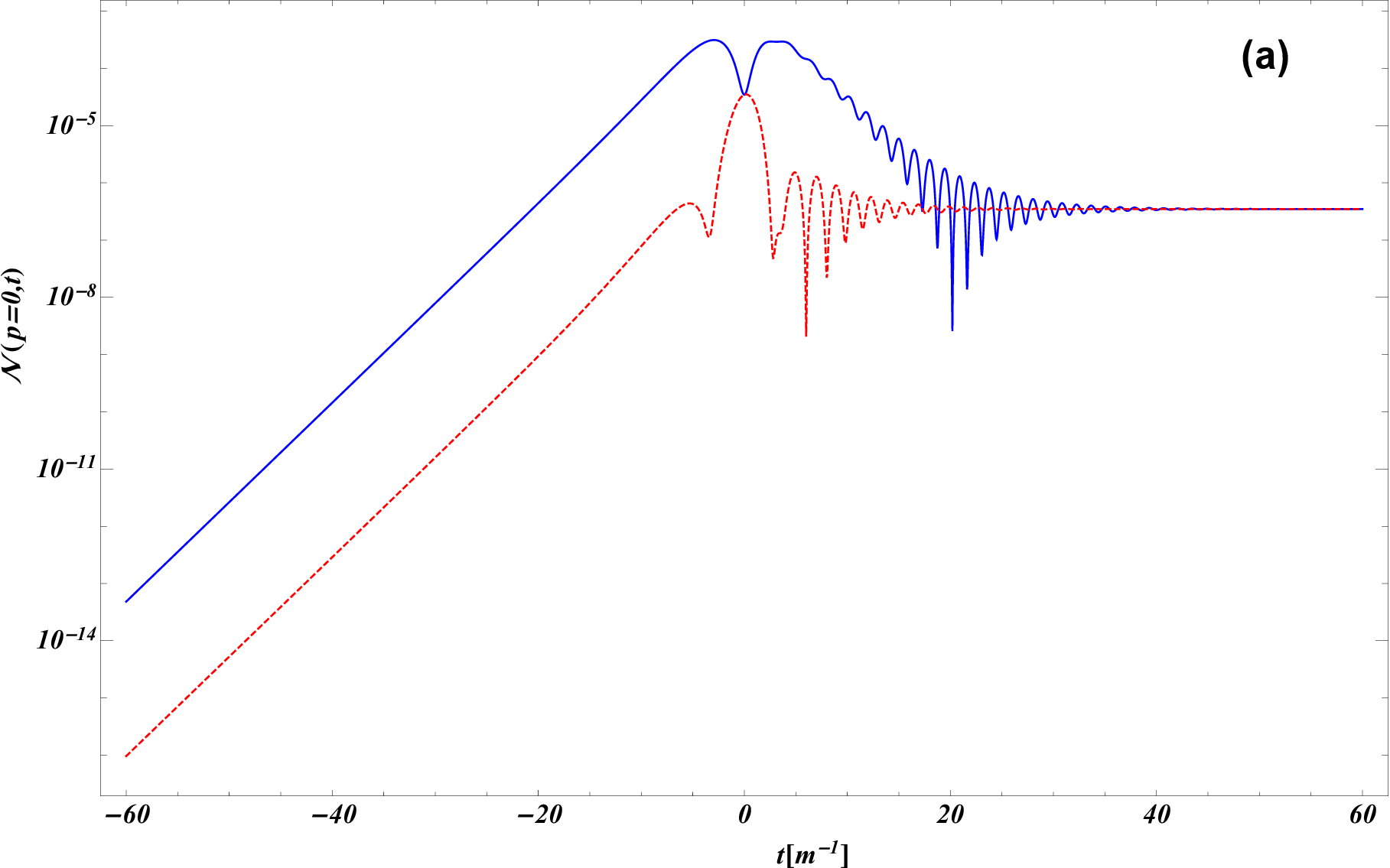}
     \includegraphics[width=0.395\linewidth]{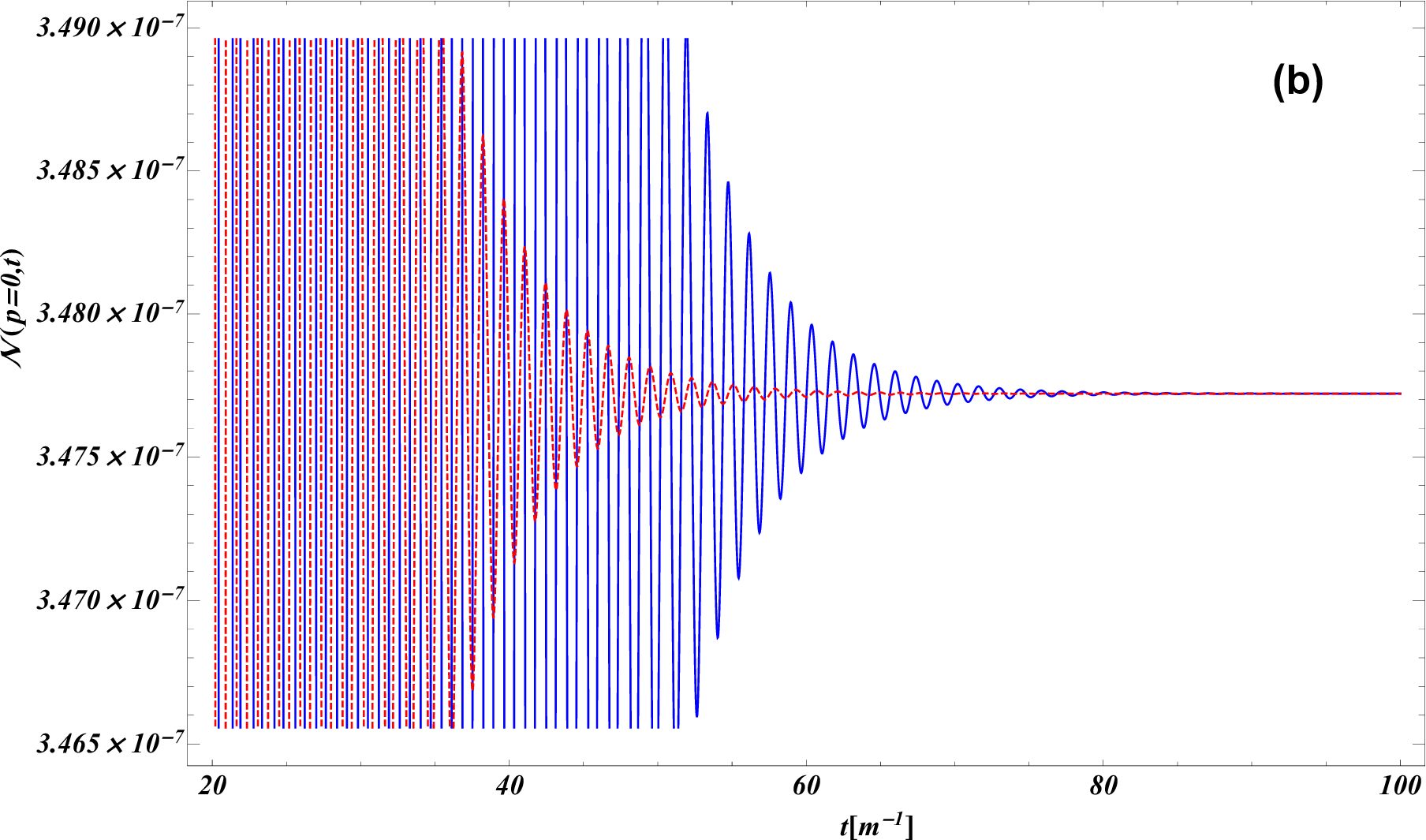}
    \caption{Evolution of $\mathcal{N}(p,t)$  for two  basis choices. The blue curve represent first  choice  of adiabatic frequency $\mathcal{W}(p,t) = \omega(p,t)$ and $\mathcal{Y}(p,t)= 0$ and dashed red curve for second choice  $\mathcal{W}(p,t) = \omega(p,t)$ and $\mathcal{Y}(p,t) = \frac{\dot{\omega}(p,t)}{ 2 \omega(p,t)^2}$ .The field parameters are  $E_0=0.2 E_c$ and $ \tau =10 [m^{-1}].$ The momentum is considered to be zero, and all the units are taken in the electron mass unit.}
    \label{N_twochoices}
\end{figure}

\section{Adiabatic Basis Dependence and Polarization Current  }
\label{invariance}
As discussed earlier in section \ref{sec_N(t)}, the choice of reference basis functions plays a crucial role in defining the time-dependent particle number \( \mathcal{N}(p,t) \) and understanding the dynamics of pair creation using Bogoliubov coefficients. At intermediate times (non-asymptotic), these coefficients yield ambiguous values, depending on the choice of adiabatic functions \( \mathcal{W}(p,t) \) and \( \mathcal{Y}(p,t) \).  

To illustrate how the time-dependent particle number depends on the reference basis, we present the temporal evolution of \( \mathcal{N}(p,t) \) for standard choices of \( \mathcal{W}(p,t) \) and \( \mathcal{Y}(p,t) \), which originate from the WKB approximation, where \( \mathcal{W}(p,t) = \omega(p,t) \). Two commonly used choices include:  
Choice 1: \( \mathcal{W}(p,t) = \omega(p,t) \), \( \mathcal{Y}(p,t) = 0 \) (as used in \cite{Birrell}).  
\newline
Choice 2: \( \mathcal{W}(p,t) = \omega(p,t) \), \( \mathcal{Y}(p,t) = \frac{\dot{\omega}(p,t)}{2 \omega(p,t)^2} \) (as used in \cite{Schmidt:1998vi}).  
\par
Figure \ref{N_twochoices} compares \( \mathcal{N}(p,t) \) for these two choices. At early times (\( t < 0 \)), significant differences emerge, with the blue curve increasing more rapidly than the red curve. During the interaction period (\( -10 < t < 20 \)), the blue curve exhibits stronger oscillations, whereas the red curve follows a smoother trend. At late times (\( t > 20 \)), both curves stabilize, approaching their asymptotic values. However, oscillations persist longer in the blue curve, whereas the red curve converges more smoothly.  

The second panel magnifies the late-time behavior (\( t > 20 \)), highlighting these differences. The persistent oscillations in the blue curve suggest that the choice without \( \mathcal{Y}(p,t)\) introduces additional transient effects, possibly due to incomplete decoupling of positive and negative frequency modes. The asymptotic values of \( \mathcal{N}(p,t) \) should be compared to determine which basis provides a more physically meaningful description over time. 

\begin{figure}
    \centering
    \includegraphics[width=0.5\linewidth]{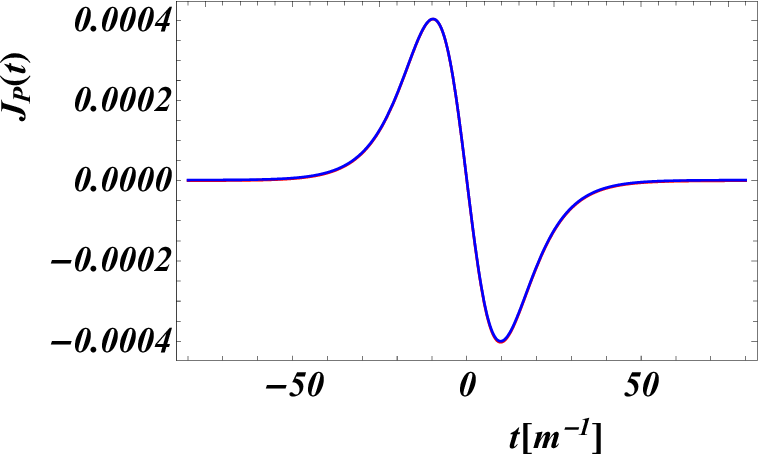}
    \caption{The time evolution of the vacuum current polarization current, $ J_P(t)$, defined with respect to two different adiabatic bases. The blue curve represent first  choice  of adiabatic frequency $\mathcal{W}(p,t) = \omega(p,t)$ and $\mathcal{Y}(p,t)= 0$ and dashed red curve for second choice  $\mathcal{W}(p,t) = \omega(p,t)$ and $\mathcal{Y}(p,t) = \frac{\dot{\omega}(p,t)}{ 2 \omega(p,t)^2}$ .The field parameters are  $E_0=0.2 E_c$ and $ \tau =10 [m^{-1}].$ The momentum is considered to be zero, and all the units are taken in the electron mass unit.}
    \label{Jpol_basis}
\end{figure}

To further investigate the physical significance of these choices, we examine the polarization current \( J_P(t) \). If \( J_P(t) \) is a true physical observable, it should remain invariant across different basis choices. We plot its temporal evolution for both reference basis functions and compare the results.  

A key aspect of vacuum dynamics during pair production is identifying quantities that are unique and well-defined at all times. The polarization current serves as one such candidate. By analyzing its behavior under different basis choices—similar to the approach used for \( \mathcal{N}(p,t) \)—we assess whether it provides a consistent physical interpretation beyond the asymptotic regime.  
The overlap of \( J_P(t) \) from the two basis choices demonstrates that the polarization current remains unique and retains clear physical meaning at all times (see Fig.\ref{Jpol_basis}). Unlike \( \mathcal{N}(p,t) \), which depends on the reference basis at intermediate times, \( J_P(t) \) is an observable quantity and, at least in principle, measurable. This suggests that polarization current may provide a more robust characterization of vacuum dynamics in pair production.
\section{Conclusion}
In this work, we studied vacuum pair creation in the presence of a time-dependent Sauter pulse field using the canonical quantization of matter fields. We derived generalized expressions for the time-dependent particle number and correlation function by allowing arbitrary mode selections via Bogoliubov transformations.
Our analysis focused on the evolution of the particle number and vacuum polarization effects, examining how different field strengths and pulse durations influence pair creation. 
The  real part of the correlation function  $u(p,t)$ gives vacuum polarization, where as the imaginary part  $v(p,t)$  represent the corresponding counter term,were computed using exact mode function solutions. We found that both functions remain negligible at early times, exhibit a strong transient peak at $t=0$, and display oscillatory behavior at late times due to quantum interference effects. Notably, $v(p,t)$ exhibits an antisymmetric structure, while $u(p,t)$ follows a peak-like evolution, highlighting the interplay between polarization and depolarization mechanisms.
The external field strength $E_0$ significantly impacts the magnitude and frequency of oscillations, with stronger fields leading to more pronounced responses. Additionally, the pulse duration $\tau$ modulates the peak values and the persistence of oscillations, revealing the temporal characteristics of vacuum fluctuations. At asymptotic times, the oscillations stabilize, marking a transition to a final equilibrium state where real particle formation dominates.
We also investigated the dynamical properties of the vacuum polarization current $J_P(t)$, which depends on the rate of particle creation and is directly related to the vacuum polarization function $u(p,t)$. The integrand of $J_P(t)$ exhibits a peak around the field maximum, followed by oscillatory behavior. A shorter pulse results in stronger oscillations and a broader frequency response, whereas longer pulses lead to smoother variations. The field strength  determines the peak magnitude and oscillation characteristics, with higher values inducing sharper transitions.
Furthermore, we examined the dependence of the time-dependent particle number $\mathcal{N}(p,t)$ on the choice of adiabatic basis functions. Our results show that $\mathcal{N}(p,t)$ varies significantly at intermediate times depending on the basis choice, whereas the polarization current $J_P(t)$ remains invariant, confirming its robustness as a physically meaningful observable.
Overall, our findings provide deeper insight into vacuum polarization effects and the interplay between particle creation and annihilation in strong-field QED. The results emphasize the importance of polarization current as a stable observable in describing the dynamics of pair production processes. 
\section{Acknowledgments}

Deepak acknowledges the financial assistance provided by the Raja Ramanna Center for Advanced Technology (RRCAT) and the Homi Bhabha National Institute (HBNI) for carrying out this research work.


\end{document}